\documentclass[journal,twoside]{IEEEtran}
\usepackage{graphicx,cite,amssymb,amsmath,psfrag,subfigure}
\usepackage{graphicx}
\usepackage{subfigure}
\usepackage{mathrsfs}
\usepackage[displaymath,mathlines]{lineno}
\usepackage{color}
\usepackage{tabulary}
\usepackage{multirow}
\usepackage{cases}
\usepackage{stfloats}
\usepackage{url}
\usepackage[T1]{fontenc}

\newtheorem{remark}{Remark}
\newtheorem{algorithm}{Algorithm}

\DeclareMathOperator{\E}{\mathsf{E}}

\DeclareMathOperator{\var}{\mathsf{Var}}
\DeclareMathOperator{\Pro}{\mathsf{Pr}}

\newcommand{\I}{{\tt I}}
\newcommand{\h}{{\tt h}}

\newcommand{\EX}[1]{\mathsf{E}\left\{{#1}\right\}}

\newcommand{\varx}[1]{\var\left\{{#1}\right\}}
\newcommand{\Prx}[1]{\Pro\left\{{#1}\right\}}

\newcommand{\CG}[2]{\mathcal{CN}\left({#1},{#2}\right)}

\newcommand{\B}[1]{{\mathbf{#1}}}

\newcommand{\Pu}{\rho_{\mathrm{u}}}

\newcommand{\Pd}{\rho_{\mathrm{d}}}

\newcommand{\tauc}{\tau_\mathrm{c}}
\newcommand{\taup}{\tau_\mathrm{u,p}}
\newcommand{\taud}{\tau_\mathrm{d}}
\newcommand{\taudp}{\tau_\mathrm{d,p}}

\newcommand{\snorm}[1]{{ \left\Vert #1 \right\Vert^2 }}

\newcounter{eqncnt}

\newcounter{eqnback}

\begin{document}
\title{
    No Downlink Pilots are Needed in TDD Massive MIMO}
\author{
        Hien Quoc Ngo, \emph{Member, IEEE}, and
        Erik G. Larsson, \emph{Fellow, IEEE}
\thanks{Manuscript received May 09, 2016; revised September 07, 2016 and October 28, 2016; accepted
November 08, 2016. The associate editor coordinating the review of
this paper and approving it for publication was Dr.~Jun Zhang.
This work was supported  in part by the Swedish Research Council
(VR), the Swedish Foundation for Strategic Research (SSF), and
ELLIIT. Part of this work was presented at the 2015 IEEE International Conference on Acoustics, Speech and Signal Processing (ICASSP) \cite{NL:15:ICASSP}.}
\thanks{
        H.~Q.\ Ngo and E.~G. Larsson are with the Department of Electrical
    Engineering (ISY), Link\"{o}ping University, 581 83 Link\"{o}ping,
    Sweden
        (Email: hien.ngo@liu.se; egl@isy.liu.se).  H.~Q.\ Ngo is also with the School of Electronics, Electrical Engineering and Computer Science, Queen's University Belfast, Belfast BT3 9DT, U.K.
}

\thanks{Digital Object Identifier xxx/xxx}
 }

\markboth{IEEE Transactions on Wireless Communications, Vol. XX, No. X, XXX
2016}{Ngo \textit{\MakeLowercase{et al.}}: No Downlink Pilots are Needed in TDD Massive MIMO}

\maketitle

\begin{abstract}
We consider the Massive Multiple-Input Multiple-Output downlink
with maximum-ratio and zero-forcing processing and time-division
duplex operation. To decode, the users must know their
instantaneous effective channel gain.  Conventionally, it is assumed
that by virtue of channel hardening, this instantaneous gain is close
to its average and hence that users can rely on knowledge of that
average (also known as statistical channel information). However, in
some propagation environments, such as keyhole channels, channel
hardening does not hold.

We propose a blind algorithm to estimate the effective channel gain at
each user, that does not require any downlink pilots. We derive a
capacity lower bound of each user for our proposed scheme, applicable
to any propagation channel. Compared to the case of no downlink pilots
(relying on channel hardening), and compared to training-based
estimation using downlink pilots, our blind algorithm performs
significantly better. The difference is especially pronounced in
environments that do not offer channel hardening.
\end{abstract}


\section{Introduction} \label{Sec:Introduction}
\label{sec:intro}

\IEEEPARstart{I}{n} Massive Multiple-Input Multiple-Output (MIMO),
the base station (BS) is equipped with a large antenna array (with
hundreds of antennas) that simultaneously serves many (tens or
more of) users. It is a key, scalable technology for next
generations of wireless networks, due to its promised huge energy
efficiency and spectral efficiency
\cite{NLM:13:ACCCC,Eri:13:MCOM,LLSAZ:14:JSTSP,BL:16:VTM,GERT:15:WCOM,ZJWZM:14:SSP}.
In Massive MIMO, time-division duplex (TDD) operation is
preferable, because the amount of pilot resources required does
not depend on the number of BS antennas. With TDD, the BS obtains
the channel state information (CSI) through uplink training.  This
CSI is used to detect the signals transmitted from users in the
uplink. On downlink, owing to the reciprocity of propagation, CSI
acquired at the BS is used for precoding.  Each user receives an
effective (scalar) channel gain multiplied by the desired symbol,
plus interference and noise. To coherently detect the desired
symbol, each user should know its effective channel gain.

Conventionally, each user is assumed to approximate its
instantaneous channel gain by its mean
\cite{JAMV:11:WCOM,YM:13:JSAC,HBD:13:JSAC}. This is known to work
well in Rayleigh fading. Since Rayleigh fading channels harden
when the number of BS antennas is large (the effective channel
gains become nearly deterministic), the effective channel gain is
close to its mean. Thus, using the mean of this gain for signal
detection works very well. This way, downlink pilots are avoided
and users only need to know the channel statistics. However, for
small or moderate numbers of antennas, the gain may still deviate
significantly from its mean.  Also, in propagation environments
where the channel does not harden, using the mean of the effective
gain as substitute for its true value may result in poor
performance even with large numbers of antennas.

The users may estimate their effective channel gain by using downlink pilots,
see \cite{NLM:13:ACCCC} for single-cell systems and \cite{ZZYJL:15:VT}
for multi-cell systems.  Effectively, these downlink pilots are
orthogonal between the users and beamformed along with the downlink
data.  The users may use,  for
example, linear minimum mean-square error (MMSE) techniques for the
estimation of this gain.  The
downlink rates of multi-cell systems for maximum-ratio (MR) and zero-forcing (ZF) precoders  with and without downlink pilots were analyzed in \cite{KM:15:COM}.
The effect of using outdated gain estimates at the users was
investigated in \cite{KMC:15:ICC}. Compared
with the case when the users rely on statistical channel knowledge,
the downlink-pilot based schemes improve the system performance in
low-mobility environments (where the coherence interval is
long). However, in high-mobility environments, they do not work well,
owing to the large requirement of downlink training resources; this
required overhead is proportional to the number of multiplexed users.
A better way of estimating  the effective channel gain, which requires less  resources than the transmission of downlink pilots does, would be desirable.

Inspired by the above discussion, in this paper, we consider the
Massive MIMO downlink with TDD operation. The BS acquires  channel
state  information  through  the  reception  of  uplink  pilot
signals transmitted by the users -- in the conventional manner,
and when transmitting data to the users, it applies
  MR or ZF processing with slow time-scale power control.
  For this system, we propose a simple blind    method for the estimation of the effective gain,
 that each user  should independently perform, and which
  does not require any downlink pilots.
 Our proposed method   exploits the asymptotic properties of the received data in each  coherence interval.
 Our specific contributions
are:
\begin{itemize}
\item We give a formal definition of channel hardening, and an
  associated criterion that can be used to test if channel hardening
  holds. Then we examine two important propagation scenarios:
  independent Rayleigh fading, and keyhole channels. We show that
  Rayleigh fading channels harden, but keyhole channels do not.

    \item We propose a blind channel estimation scheme, that each user
      applies in the downlink.  This scheme exploits the asymptotic
      properties of the sample average power of the received signal
      per coherence interval. We presented a preliminary version of
      this algorithm in \cite{NL:15:ICASSP}.

    \item We derive a rigorous capacity lower bound for Massive MIMO
      with estimated downlink channel gains.  This bound can be
      applied to any types of channels and can be used to analyze the
      performance of any downlink channel estimation method.

      \item Via numerical results we show that, in hardening
      propagation environments, the performance of our proposed blind
      scheme is comparable to the use of only statistical channel
      information (approximating the gain by its mean).  In contrast,
      in non-hardening propagation environments, our proposed scheme
      performs much better than the use of statistical channel
      information only. The results also show that our blind method
      uniformly outperforms schemes based on downlink pilots
      \cite{NLM:13:ACCCC,ZZYJL:15:VT}.
\end{itemize}

\textit{Notation:} We use boldface upper- and lower-case letters
to denote matrices and column vectors, respectively. Specific
notation and symbols used in this paper are listed as follows:

\begin{tabular}{l l}
  $()^\ast$, $()^T$, and $()^H$ & Conjugate, transpose, and transpose \\
  &  conjugate\\

  $\det\left(\cdot \right)$ and $\mathrm{Tr}\left(\cdot\right)$ & Determinant and trace of a matrix\\

  $\CG{\mathbf{0}}{\mathbf{\Sigma}}$ & Circularly symmetric complex \\
  &Gaussian  vector with zero mean \\
  & and covariance matrix $\mathbf{\Sigma}$\\

  $|\cdot|$, $\|\cdot\|$ & Absolute value, Euclidean norm\\

  $\mathsf{E}\left\{\cdot\right\}$, $\varx{\cdot}$& Expectation, variance operators\\

  $\mathop \to \limits^{P}$ & Convergence in probability\\

  $\B{I}_n$ & $n\times n$ identity matrix\\

 $\left[\B{A} \right]_k$, $\B{a}_k$ &The $k$th column of $\B{A}$.\\
\end{tabular}

\section{System Model} \label{Sec:SysModel}

We consider a single-cell Massive MIMO system with an $M$-antenna
BS and $K$ single-antenna users, where $M>K$. The channel between
the BS and the $k$th user is an $M\times 1$ channel vector,
denoted by $\B{g}_k$, and is modelled as:
\begin{align}\label{eq:sys1}
\B{g}_k = \sqrt{\beta_k}\B{h}_k,
\end{align}
where $\beta_k$ represents large-scale fading which is constant
over many coherence intervals, and $\B{h}_k$ is an $M\times 1$
small-scale fading channel vector. We assume that the elements of
$\B{h}_k$ are uncorrelated, zero-mean and unit-variance random
variables (RVs) which are not necessarily Gaussian distributed.
Furthermore, $\B{h}_k$ and $\B{h}_{k'}$ are assumed to be
independent, for $k\neq k'$. The $m$th elements of $\B{g}_k$ and
$\B{h}_k$ are denoted by $g_{k}^m$ and $h_{k}^m$, respectively.

Here, we focus on the downlink data transmission with TDD
operation. The BS uses the channel estimates obtained in the
uplink training phase, and applies MR or ZF processing to transmit
data to all  users in the same time-frequency resource.

\subsection{Uplink Training}

Let $\tauc$ be the length of the coherence interval (in symbols).
For each coherence interval, let $\taup$ be the length of uplink
training duration (in symbols). All users simultaneously send
pilot sequences of length $\taup$ symbols each to the BS. We
assume that these pilot sequences are pairwisely orthogonal. So it
is required that $\taup\geq K$. The linear MMSE estimate of
$\B{g}_k$ is given by \cite{NLM:13:TCOM}
\begin{align}\label{eq:CE1}
\hat{\B{g}}_k
    =
        \frac{\taup\Pu\beta_k}{\taup\Pu\beta_k+1}\B{g}_k +
        \frac{\sqrt{\taup\Pu}\beta_k}{\taup\Pu\beta_k+1}\B{w}_{\text{p},k},
\end{align}
where $\B{w}_{\text{p},k}\sim\CG{0}{\B{I}_M}$ independent of
$\B{g}_k$, and $\Pu$ is the transmit signal-to-noise ratio (SNR)
of each pilot symbol.

The variance of the $m$th element of $\hat{\B{g}}_k$ is given by
\begin{align}\label{eq:varCE1}
\varx{\hat{g}_{k}^m}
    =
    \EX{|\hat{g}_{k}^m|^2}
    =
    \frac{\taup\Pu\beta_k^2}{\taup\Pu\beta_k+1}
    \triangleq \gamma_k.
\end{align}

Let $\tilde{\B{g}}_{k} = \B{g}_{k} - \hat{\B{g}}_{k}$ be the channel
estimation error, and $\tilde{g}_{k}^m$ be the $m$th element of
$\tilde{\B{g}}_{k}$. Then from the properties of linear MMSE estimation,
$\tilde{g}_{k}^m$ and $\hat{g}_{k}^m$ are uncorrelated, and
\begin{align}\label{eq:varCEE1}
\varx{\tilde{g}_{k}^m}
    =
    \EX{|\tilde{g}_{k}^m|^2}
    =
    \beta_k - \gamma_k.
\end{align}
In the special case where $\B{g}_k$ is Gaussian distributed
(corresponding to Rayleigh fading channels), the linear MMSE estimator
becomes the MMSE estimator and $\tilde{g}_{k}^m$ is independent of
$\hat{g}_{k}^m$.

\subsection{Downlink Data Transmission}

Let $s_k(n)$ be the $n$th symbol intended for the $k$th user. We
assume that $\EX{\B{s}(n)\B{s}(n)^H}=\B{I}_K$, where
$\B{s}(n)\triangleq [s_1(n), \ldots, s_K(n)]^T$. With linear
processing, the $M\times 1$ precoded signal vector is
\begin{align}\label{eq:sys2}
\B{x}(n) = \sqrt{\Pd}\sum_{k=1}^K\sqrt{\eta_k}\B{a}_{k}s_{k}(n),
\end{align}
where $\{\B{a}_k\}$, $k=1, \ldots, K$, are the
precoding vectors which are functions of the channel estimate
$\hat{\B{G}}\triangleq [\hat{\B{g}}_1, \ldots, \hat{\B{g}}_K]$,
${\Pd}$ is the (normalized) average transmit power, $\{\eta_k\}$
are the power coefficients, and $\B{D}_\eta$ is a diagonal matrix
with $\{\eta_k\}$ on its diagonal. For a given $\{\B{a}_k\}$, the power control coefficients $\{\eta_k\}$ are
chosen to satisfy an average power constraint at the BS:
\begin{align}\label{eq:powerconst}
\EX{\|\B{x}(n)\|^2}\leq \Pd.
\end{align}

The signal received at the $k$th user is\footnote{Here we restrict our
  consideration to one coherence interval so that the channels remain
  constant.}
\begin{align}\label{eq:sys3}
&y_k(n) =
        \B{g}_k^H\B{x}(n) + w_k(n)\nonumber\\
    &=
        \sqrt{\Pd\eta_k}\alpha_{kk} s_k(n) + \sum_{k'\neq k}^K \sqrt{\Pd\eta_{k'}}\alpha_{kk'}s_{k'}(n) +
    w_k(n),
\end{align}
where $w_k(n)\sim\CG{0}{1}$ is additive Gaussian noise, and
$$
    \alpha_{kk'} \triangleq \B{g}_k^H\B{a}_{k'}.
$$
Then, the desired signal $s_k$ is decoded.

We consider two linear precoders: MR and
ZF processing.

\begin{itemize}
    \item MR processing: here the  precoding vectors $\{\B{a}_k\}$
are
\begin{align}\label{eq:mr1}
    \B{a}_k
    =
    \frac{\hat{\B{g}}_k}{\|\hat{\B{g}}_k\|}, \quad k=1, \ldots, K.
\end{align}

    \item ZF processing: here the precoding vectors are
\begin{align}\label{eq:zf111}
\B{a}_k=\frac{1}{\left\|\left[\hat{\B{G}}\left(\hat{\B{G}}^H\hat{\B{G}}
\right)^{-1}\right]_k
\right\|}\left[\hat{\B{G}}\left(\hat{\B{G}}^H\hat{\B{G}}
\right)^{-1}\right]_k,
\end{align}
for $k=1, \ldots, K$.
\end{itemize}
With the precoding vectors given in \eqref{eq:mr1} and
\eqref{eq:zf111}, the power constraint \eqref{eq:powerconst}
becomes
\begin{align}\label{eq:powerconstbc}
\sum_{k=1}^K \eta_k \leq 1.
\end{align}

\section{Preliminaries of Channel Hardening}\label{sec:CH}

One motivation of this work is that Massive MIMO channels may not
always harden.  In this section we discuss the channel hardening
phenomena. We specifically study channel hardening for independent
Rayleigh fading and for keyhole channels.

Channel hardening is a phenomenon where the norms of the channel
vectors $\{\B{g}_k\}$, $k=1,\ldots, K$, fluctuate only little. We say
that the propagation offers \emph{channel hardening} if
\begin{align}\label{eq:defch1}
\frac{\|\B{g}_k\|^2}{\EX{\|\B{g}_k\|^2}}  ~ \mathop
{\to}\limits^{P} ~ 1, \quad \text{as} ~ M\to\infty, \quad k=1,
\ldots, K.
\end{align}

\subsection{Advantages of Channel Hardening}
If the BS and the users know the channel $\B{G}$ perfectly, the
channel is deterministic and its sum-capacity is given by
\cite{VT:03:IT}
\begin{align}\label{eq:macrate1a}
C  = \max_{\eta_k\geq 0, \sum_{k=1}^K\eta_k \leq 1}\log_2
\det\left(\B{I}_M + \Pd\B{G} \B{D}_{\eta} \B{G}^H \right),
\end{align}
where $\B{D}_\eta$ is the diagonal matrix whose $k$th diagonal
element is the power control coefficient $\eta_k$.

In Massive MIMO, for most propagation environments, we have
asymptotically favorable propagation \cite{NLM:14:Eusipco}, i.e.
$\frac{\B{g}_k^H\B{g}_{k'}}{M}\to 0$, as $M\to\infty$, for $k\neq
k'$. In addition, if the channel hardens, i.e.,  $\frac{\|\B{g}_k\|^2}{M}\to\EX{\snorm{\B{g}_k}}=\beta_k$, as $M\to\infty$,\footnote{ Note that \emph{favorable propagation} and \emph{channel hardening} are two different properties of the channels. Favorable propagation,  $\frac{1}{M}{\B{g}_k^H\B{g}_{k'}}\to 0$ as $M\to\infty$, does not imply  hardening,  $\frac{1}{M}{\|\B{g}_k\|^2}\to\beta_k$. One example of the contrary is the keyhole channel in Section~\ref{sec:keyh}.} then we have, for fixed $K$,
 \begin{align}\label{eq:CFPCH1a}
&C -  \max_{\eta_k\geq 0, \sum_{k=1}^K\eta_k \leq 1}
\sum_{k=1}^K \log_2 \left(   1 + \Pd \eta_k \beta_k M \right)\nonumber\\
&=C - \!\!\!
\max_{\eta_k\geq 0, \sum_{k=1}^K\eta_k \leq 1}\!\!\!\!\!\!\log_2
\det\!\!\left(\B{I}_K \!+\! \Pd  \B{D}_{\eta} M  \left[ \begin{array} {ccc}
\beta_1 & \cdots & 0 \\
\vdots &  \ddots & \vdots \\
0 &\cdots & \beta_K \\
\end{array} \right] \right)\nonumber\\
&=\!\!\!\!\!\!
\max_{\eta_k\geq 0, \sum_{k=1}^K\eta_k \leq 1}\!\!\!\!\!\!\log_2
\det\!\!\left(  \left[ \begin{array} {ccc}
\frac{1+\Pd \eta_1\|\B{g}_1\|^2}{1+\Pd \eta_1 \beta_1 M} & \cdots & \frac{\Pd\eta_1\B{g}_1^H\B{g}_K}{1+\Pd\eta_K\beta_K M} \\
\vdots &  \ddots & \vdots \\
\frac{\Pd\eta_K\B{g}_K^H\B{g}_1}{1+\Pd\eta_1\beta_1 M} &\cdots & \frac{1+\Pd \eta_K\|\B{g}_K\|^2}{1+\Pd \eta_K \beta_K M} \\
\end{array} \right] \right)\
\nonumber\\
&\to 0, \quad \text{as} ~ M\to\infty.
\end{align}
In \eqref{eq:CFPCH1a} we have used the facts that
 \begin{align*}
\frac{1+\Pd \eta_k\|\B{g}_k\|^2}{1+\Pd \eta_k \beta_k M}
=
\frac{\frac{1}{M}+\Pd \eta_k\frac{\|\B{g}_k\|^2}{M}}{\frac{1}{M}+\Pd \eta_k \beta_k }
\to 1, ~ \text{as} ~ M\to\infty,
\end{align*}
and for $k\neq k'$,
 \begin{align*}
\frac{\Pd\eta_k\B{g}_k^H\B{g}_{k'}}{1+\Pd\eta_{k'}\beta_{k'}M}
=
\frac{\Pd\eta_k\B{g}_k^H\B{g}_{k'}/M}{1/M+\Pd\eta_{k'}\beta_{k'}}
\to 0, \quad \text{as} ~ M\to\infty.
\end{align*}
The limit in   (\ref{eq:CFPCH1a})  implies that if the channel
hardens, the sum-capacity (\ref{eq:macrate1a}) can be
approximated for $M\gg K$ as:
 \begin{align}\label{eq:CFPCH2}
C  \approx \max_{\eta_k\geq 0, \sum_{k=1}^K\eta_k \leq 1}
\sum_{k=1}^K \log_2 \left(   1 + \Pd \eta_k \beta_k M \right),
\end{align}
which does not depend on the small-scale fading. As a consequence, the
system scheduling, power allocation, and interference management can
be done over the large-scale fading time scale instead of the
small-scale fading time scale. Therefore, the overhead for these
system designs is significantly reduced.


Another important advantage is: if the channel hardens, then we do not
need instantaneous CSI at the receiver to detect the transmitted
signals. What the receiver needs is only the statistical knowledge of
the channel gains. This reduces the resources (power and training
duration) required for channel estimation.  More precisely, consider
the signal received at the $k$th user given in \eqref{eq:sys3}. The
$k$th user wants to detect $s_k$ from $y_k$. For this purpose, it
needs to know the effective channel gain $\alpha_{kk}$. If the channel
hardens, then $\alpha_{kk} \approx \EX{\alpha_{kk}}$. Therefore, we
can use the statistical properties of the channel, i.e.,
$\EX{\alpha_{kk}}$ is a good estimate of $\alpha_{kk}$ when detecting
$s_k$. This assumption is widely made in the Massive MIMO literature
\cite{JAMV:11:WCOM,YM:13:JSAC,HBD:13:JSAC} and circumvents the need
for downlink channel estimation.

\subsection{Measure of Channel Hardening}

We next state a simple criterion, based on  the Chebyshev inequality, to check whether the channel
hardens or not. A similar method was discussed in \cite{LCC:15:WCOM}. From   Chebyshev's inequality, we have
\begin{align}\label{eq:MCH1}
    &\Prx{\left|\frac{\snorm{\B{g}_k}}{\EX{\snorm{ \B{g}_k }}} -1 \right|^2 \leq \epsilon}\nonumber\\
    &=
    1-\Prx{\left|\frac{\snorm{\B{g}_k}}{\EX{\snorm{ \B{g}_k }}} -1 \right|^2 \geq
    \epsilon}  \nonumber\\
    &\geq 1 - \frac{1}{\epsilon}\cdot\frac{\varx{\snorm{\B{g}_k}}
    }{\left(\EX{\snorm{\B{g}_k}}\right)^2}, \quad \text{for any $\epsilon \geq
    0$}.
\end{align}
Clearly, if
\begin{align}\label{eq:MCH1bb}
\frac{\varx{\snorm{\B{g}_k}}
    }{\left(\EX{\snorm{\B{g}_k}}\right)^2} \to 0, \quad \text{as} ~ M\to\infty,
\end{align}
 we have channel
    hardening.
In contrast, \eqref{eq:defch1} implies
$$\frac{\varx{\snorm{\B{g}_k}} }{\left(\EX{\snorm{\B{g}_k}}\right)^2}
\to 0, \quad \text{as} ~ M\to\infty,$$ so if \eqref{eq:MCH1bb} does
not hold, then the channel does not harden.  Therefore, we can use
$\frac{\varx{\snorm{\B{g}_k}} }{\left(\EX{\snorm{\B{g}_k}}\right)^2}$
to determine if channel hardening holds for a particular propagation
environment.

\subsection{Independent Rayleigh Fading and Keyhole Channels}\label{sec:Channels}

In this section, we study the channel hardening property of two
particular channel models: Rayleigh fading and keyhole channels.

\subsubsection{Independent Rayleigh Fading Channels}

Consider the channel model \eqref{eq:sys1} where $\{h_k^m\}$ (the
elements of $\B{h}_k$) are i.i.d.\ $\CG{0}{1}$ RVs. Independent
Rayleigh fading channels occur in a dense, isotropic scattering
environment \cite{moustakas2000communication}. By using the
identity $\EX{\|\B{g}_k\|^4}=\beta_k^2(M+1)M$ \cite{TV:04:FTCIT},
we obtain
 \begin{align}\label{eq:iidrayfp1}
\frac{\varx{\snorm{\B{g}_k}}
    }{\left(\EX{\snorm{\B{g}_k}}\right)^2}
    &=\frac{1}{\beta_k^2 M^2} \EX{\|\B{g}_k\|^4}-1\nonumber\\
    &=\frac{1}{M} \to 0, \quad
    M\to\infty.
\end{align}
Therefore, we have channel hardening.

\subsubsection{Keyhole Channels}\label{sec:keyh}

\begin{figure}[t!]
\centerline{\includegraphics[width=0.5\textwidth]{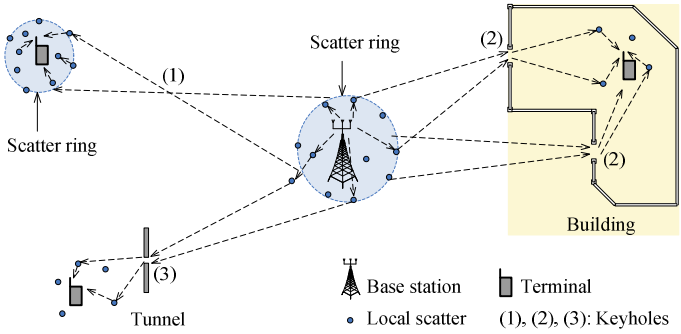}}
\caption[KH]{Examples of keyhole channels: (1)---keyhole effects
  occur when the distance between transmitter and receiver is
  large. The transmitter and the receiver have their own local
  scatters which yield locally uncorrelated fading. However, the
  scatter rings are much smaller than the distance between them, the
  channel becomes low rank, and hence keyhole effects occur
  \cite{GBGP:02:COM}; (2)---the receiver is located inside a building,
  the only way for the radio wave to propagation from the transmitter
  to the receiver is to go through several narrow holes which can be
  considered as keyholes; and (3)---the transmitter and the receiver
  are separated by a tunnel.\label{fig:KH_1}}
\end{figure}

A keyhole channel (or double scattering channel) appears in
scenarios with rich scattering around the transmitter and receiver,
and where there is a low-rank connection between the two scattering
environments. The keyhole effect can occur when the radio wave goes
through tunnels,  corridors, or when the distance between the
transmitter and receiver is large.  Figure~\ref{fig:KH_1} shows some
examples where the keyhole effect occurs in practice. This channel
model has been validated both in theory and by practical experiments
\cite{ATM:06:WCOM,LJGM:10:SP,ZJWM:11:WCOM,LL:IT:11}. Under keyhole
effects, the channel vector $\B{g}_k$ in \eqref{eq:sys1} is modelled
as \cite{LJGM:10:SP}:
\begin{align}\label{eq:ds 1}
    \B{g}_k = \sqrt{\beta_k}\sum_{j=1}^{n_k} c_j^{(k)} a_j^{(k)}
    \B{b}_j^{(k)},
\end{align}
where $n_k$ is the number of effective keyholes, $a_j^{(k)}$ is
the random channel gain from the $k$th user to the $j$th keyhole,
$\B{b}_j^{(k)}\in \mathbb{C}^{M\times 1}$ is the random channel
vector between the $j$th keyhole associated with the $k$th user
and the BS, and $c_j^{(k)}$ represents the deterministic complex
gain of the $j$th keyhole associated with the $k$th user. The
elements of $\B{b}_j^{(k)}$ and $a_j^{(k)}$ are
i.i.d. $\CG{0}{1}$ RVs. Furthermore, the gains $\{c_j^{(k)}\}$ are
normalized such that $\EX{|{g}_k^m |^2}=\beta_k$. Therefore,
\begin{align}\label{eq:ds 2}
    \sum_{i=1}^{n_k} \left|c_i^{(k)}\right|^2=1.
\end{align}
When $n_k=1$, we have a degenerate keyhole (single-keyhole)
channel. Conversely, when $n_k\to\infty$,
under the additional assumptions that
 $c_i^{(k)}\neq 0$ for finite $n_k$ and $c_i^{(k)}\to 0$ as $n_k\to\infty$,
we
obtain an i.i.d.  Rayleigh fading channel.

We assume that different users have different sets of keyholes.  This
assumption is reasonable if the users are located at random in a large
area,  as illustrated in Figure~\ref{fig:KH_1}. Then from the derivations
in Appendix~\ref{sec:AppeKU1}, we obtain
\begin{align} \label{eq:ds 5b}
\frac{\varx{\snorm{\B{g}_k}}
    }{\left(\EX{\snorm{\B{g}_k}}\right)^2}
    &=
    \left(1+\frac{1}{M}\right) \sum_{i=1}^{n_k} \left|
    c_i^{(k)}\right|^4 + \frac{1}{M}\nonumber\\  
    &\to \sum_{i=1}^{n_k} \left|c_i^{(k)}\right|^4 \neq 0, \quad
    M\to\infty.
\end{align}

Consequently, the  keyhole channels do not harden.  In
addition, since $\left| c_i^{(k)}\right|^2\leq 1$, we have
\begin{align}\label{eq:ds 6}
\frac{\varx{\snorm{\B{g}_k}}
    }{\left(\EX{\snorm{\B{g}_k}}\right)^2}
    &\leq
    \left(1+\frac{1}{M}\right) \sum_{i=1}^{n_k} \left|
    c_i^{(k)}\right|^2 + \frac{1}{M}.
\end{align}
Using \eqref{eq:ds 2}, \eqref{eq:ds 6} becomes
\begin{align}\label{eq:ds 7}
\frac{\varx{\snorm{\B{g}_k}}
    }{\left(\EX{\snorm{\B{g}_k}}\right)^2}
    &\leq
    1+ \frac{2}{M},
\end{align}
where the right hand side corresponds to the case of single-keyhole
channels ($n_k=1$). This implies that  a
single-keyhole channel represents the worst case in the sense that
then the channel gain fluctuates the most.

\section{Proposed Downlink Blind Channel Estimation Technique}

The $k$th user should know the effective channel gain
$\alpha_{kk}$ to coherently detect the transmitted signal $s_k$
from $y_k$ in \eqref{eq:sys3}. Most previous works on Massive MIMO
assume that $\EX{\alpha_{kk}}$ is used in lieu of the true
$\alpha_{kk}$ when detecting $s_k$.  The reason behind this is
that if the channel is subject to independent Rayleigh fading (the
scenario considered in most previous Massive MIMO works), it
hardens when the number of BS antennas is large, and hence
$\alpha_{kk} \approx \EX{\alpha_{kk}}$;  $\EX{\alpha_{kk}}$ is
then a good estimate of $\alpha_{kk}$. However, as seen in
Section~\ref{sec:CH}, under other propagation models the channel
may not always harden when $M\to\infty$ and then, using
$\EX{\alpha_{kk}}$ as the true effective channel $\alpha_{kk}$ to
detect $s_k$ may result in poor performance.

For the reasons explained, it is desirable that the users estimate
their effective channels.  One way to do this is to have the BS
transmit beamformed downlink pilots \cite{NLM:13:ACCCC}. Then at
least $K$ downlink pilot symbols are required.  This can
significantly reduce the spectral efficiency. For example, suppose
$M=200$ antennas serve $K=50$ users, in a coherence interval of
length $200$ symbols. If half of the coherence interval is used
for the downlink, then with the downlink beamforming training of
\cite{NLM:13:ACCCC}, we need to spend at least $50$ symbols for
sending pilots. As a result, less than $50$ of the $100$ downlink
symbols are used for payload in each coherence interval, and the
insertion of the downlink pilots reduces the overall (uplink +
downlink) spectral efficiency by a factor of $1/4$.

In what follows, we propose a blind channel estimation method
which does not require any downlink pilots.

\subsection{Downlink Blind Channel Estimation Algorithm}

We next describe our downlink blind channel estimation algorithm, a
refined version of the scheme in \cite{NL:15:ICASSP}.  Consider the
sample average power of the received signal at the $k$th user per
coherence interval:
\begin{align}\label{eq:al1}
\xi_k \triangleq \frac{|y_k(1)|^2+ |y_k(2)|^2+ \ldots+
|y_k(\taud)|^2}{\taud},
\end{align}
where $y_k(n)$ is the $n$th sample received at the $k$th user and
$\taud$ is the number of symbols per coherence interval spent on
downlink transmission. From \eqref{eq:sys3}, and by using the law of large numbers, we have, as
$\taud\to\infty$,
\begin{align}\label{eq:po1}
\xi_k   -  \left(\Pd\eta_k\left|\alpha_{kk}\right|^2 +
\sum_{k'\neq k}^K \Pd\eta_{k'}\left|\alpha_{kk'}\right|^2 +
1\right) ~ \mathop \to \limits^{P} ~ 0.
\end{align}

Since $\sum_{k'\neq k}^K \Pd\eta_{k'}\left|\alpha_{kk'}\right|^2$ is a
sum of many terms, it can be approximated by its mean (this follows from the law of large numbers).  As a
consequence, when $K$, and $\taud$ are large, $\xi_k$ in
\eqref{eq:al1} can be approximated as follows:
\begin{align}\label{eq:appro1}
\xi_k  \approx \Pd\eta_k |\alpha_{kk}|^2 + \Pd\EX{\sum_{k'\neq
k}^K\eta_{k'}|\alpha_{kk'}|^2} + 1.
\end{align}
Furthermore, the approximation \eqref{eq:appro1} is still good even if $K$ is small. The reason is that  when $K$ is small, with high probability the term $\sum_{k'\neq
k}^K\eta_{k'}|\alpha_{kk'}|^2$ is much smaller than $\eta_k |\alpha_{kk}|^2$,  since with high probability $|\alpha_{kk'}|^2\ll |\alpha_{kk}|^2$. As a result, $\sum_{k'\neq
k}^K\eta_{k'}|\alpha_{kk'}|^2$ can be approximated by its mean even for small $K$. (In fact, in the special case of $K=1$, this sum is zero.)

Equation \eqref{eq:appro1} enables us to estimate the amplitude of
the effective channel gain $\alpha_{kk}$ using the received
samples via $\xi_k$ as follows:
\begin{align}\label{eq:appro2}
\widehat{|\alpha_{kk}|}
    =
\sqrt{\frac{\xi_k- 1 - \Pd\EX{\sum_{k'\neq
k}^K\eta_{k'}|\alpha_{kk'}|^2}}{\Pd\eta_{k}}}.
\end{align}
In case the argument of the square root is non-positive, we set
the estimate $|\alpha_{kk}|$ equal to $\EX{|\alpha_{kk}|}$.

For completeness, the $k$th user also needs to estimate the phase of
$\alpha_{kk}$. When $M$ is large, with high probability, the real part
of $\alpha_{kk}$ is much larger than the imaginary part of
$\alpha_{kk}$. Thus, the phase of $\alpha_{kk}$ is very small and can
be set to zero. Based on that observation, we propose to treat the
estimate of $|\alpha_{kk}|$ as the estimate of the true $\alpha_{kk}$:
$\hat{\alpha}_{kk} = \widehat{|\alpha_{kk}|}$

The algorithm for estimating the downlink effective channel gain
$\alpha_{kk}$ is summarized as follows:

\vspace{0.4cm}

\noindent\fbox{%
    \parbox{0.475\textwidth}{

~

\begin{algorithm} \label{algo:blce}(Blind downlink channel estimation
method)
\begin{itemize}
  \item[1.] For each coherence interval, using a data block of $\taud$ samples $y_k(n)$, compute $\xi_k$ according to \eqref{eq:al1}.

 \item[2.] The $k$th user acquires  $\eta_k$ and $\EX{\sum_{k'\neq k}^K\eta_{k'}|\alpha_{kk'}|^2}$. See Remark~\ref{re:al1} for a detailed discussion on how to acquire these values.

  \item[3.] The estimate of the effective channel gain $\alpha_{kk}$
    is as
\begin{align}\label{eq:alg1}
\hspace{-0.2cm}\hat{\alpha}_{kk}
    \!=\!
    \left\{\!
\begin{array}{l}
   \sqrt{\frac{\xi_k- 1 - \Pd\EX{\sum_{k'\neq
k}^K\eta_{k'}|\alpha_{kk'}|^2}}{\Pd\eta_{k}}}, \\ \quad \text{if
$\xi_k > 1 + \Pd\EX{\sum_{k'\neq
k}^K\eta_{k'}|\alpha_{kk'}|^2}$}\\
   \EX{|\alpha_{kk}|}, \quad \text{otherwise}. \\
\end{array}%
\right.
\end{align}
\end{itemize}
\end{algorithm}
}}

~

\begin{remark}\label{re:al1}
To implement Algorithm~\ref{algo:blce}, the $k$th user has to know
$\eta_k$ and $\EX{\sum_{k'\neq k}^K\eta_{k'}|\alpha_{kk'}|^2}$. We
assume that the $k$th user knows these values. This assumption is
reasonable since these values depend only on the large-scale
fading coefficients, which stay constant over many coherence
intervals. The BS can compute these values and inform the $k$th
user about them. In addition  $\EX{\sum_{k'\neq
k}^K\eta_{k'}|\alpha_{kk'}|^2}$ can be expressed in closed form
(except for in the case of ZF processing with keyhole channels) as
follows:
\begin{align}\label{eq:mean_interference1}
&\EX{\sum_{k'\neq k}^K\eta_{k'}|\alpha_{kk'}|^2}
   \! =\left\{\!\!
\begin{array}{l}
  \sum\limits_{k'\neq k}^K\eta_{k'}\beta_k, \quad \text{for  MR,}\\\hspace{0.0cm}\text{(Rayleigh/keyhole channels)} \\
   \sum\limits_{k'\neq k}^K\eta_{k'}(\beta_k-\gamma_k), ~ \text{for ZF.}\\\hspace{0.0cm}\text{(Rayleigh channels)} \\
\end{array}%
\right.
\end{align}
Detailed derivations of \eqref{eq:mean_interference1} are
presented in Appendix~\ref{sec:app-mean_interference1}.
\end{remark}

\subsection{Asymptotic Performance Analysis} \label{Sec:PA}

In this section, we analyze the accuracy of our proposed downlink
blind channel estimation scheme when $\tauc$ and $M$ go to
infinity for two specific propagation channels: Rayleigh fading
and keyhole channels. We use the model \eqref{eq:ds 1} for keyhole
channels. When $\tauc\to\infty$, $\xi_k$ in \eqref{eq:al1} is
equal to its asymptotic value:
\begin{align}\label{eq:po1as}
\xi_k    -  \left(\Pd\eta_k\left|\alpha_{kk}\right|^2 +
\sum_{k'\neq k}^K \Pd\eta_{k'}\left|\alpha_{kk'}\right|^2 +
1\right) \to 0,
\end{align}
and hence, the channel estimate $\hat{\alpha}_{kk}$ in
\eqref{eq:alg1} becomes
\begin{align}\label{eq:asymp1}
\hat{\alpha}_{kk}
    =
    \left\{%
\begin{array}{l}
   \sqrt{|\alpha_{kk}|^2 +  \sum\limits_{k'\neq
k}^K\frac{\eta_{k'}}{\eta_k}\left(|\alpha_{kk'}|^2-\EX{|\alpha_{kk'}|^2}\right)},
\\\vspace{-0.7cm} \\ \hspace{1.2cm}\text{if $\xi_k
> 1 + \Pd\EX{\sum\limits_{k'\neq
k}^K\eta_{k'}|\alpha_{kk'}|^2}$}, \\
   \EX{|\alpha_{kk}|}, \quad \text{otherwise}. \\
\end{array}%
\right.
\end{align}

Since $\tauc\to\infty$, it is reasonable to assume that the BS can
perfectly estimate the channels in the uplink training phase,
i.e., we have
 $\hat{\B{G}}=\B{G}$. (This can be achieved by using very long uplink training
 duration.) With this assumption, $\alpha_{kk}$ is a positive real
 value. Thus, \eqref{eq:asymp1} can be rewritten as
\begin{align}\label{eq:asymp2}
\frac{\hat{\alpha}_{kk}}{\alpha_{kk}}
    =
    \left\{%
\begin{array}{l}
   \sqrt{1 +  \sum\limits_{k'\neq
k}^K\frac{\eta_{k'}}{\eta_k}\frac{|\alpha_{kk'}|^2-\EX{|\alpha_{kk'}|^2}}{\alpha_{kk}^2}},
\\\hspace{1cm} \text{if $\xi_k
> 1 + \Pd\EX{\sum\limits_{k'\neq
k}^K\eta_{k'}|\alpha_{kk'}|^2}$}, \\
   \frac{\EX{\alpha_{kk}}}{\alpha_{kk}}, \quad \text{otherwise}. \\
\end{array}%
\right.
\end{align}

\subsubsection{Maximum-Ratio Processing}

With MR processing, from \eqref{eq:mean_interference1}
and \eqref{eq:asymp2}, we have
\begin{align}\label{eq:asymp_mr1}
\frac{\hat{\alpha}_{kk}}{\alpha_{kk}}
    =
    \left\{%
    \begin{array}{l}
        \sqrt{1 +  \sum\limits_{k'\neq k}^K\frac{\eta_{k'}}{\eta_k}\frac{\left|\frac{\B{g}_k^H\B{g}_{k'}}{\left\|\B{g}_{k'}\right\|}\right|^2-\beta_k}{\left\|\B{g}_k\right\|^2}},
\\\hspace{1.9cm} \text{if $\xi_k > 1 + \Pd\sum\limits_{k'\neq k}^K\eta_{k'}\beta_k$}, \\
        \frac{\EX{\left\|\B{g}_k\right\|}}{\left\|\B{g}_k\right\|}, \quad \text{otherwise}.\\
    \end{array}%
    \right.
\end{align}

\begin{itemize}
    \item[-] \textbf{Rayleigh fading channels}: Under Rayleigh fading
    channels, $\alpha_{kk} = \left\|\B{g}_k\right\|$, and hence,
\begin{align}\label{eq:asymp_mr_rayleigh1}
    &\Prx{\xi_k > 1 + \Pd\sum_{k'\neq k}^K\eta_{k'}\beta_k}\nonumber\\
    &=
    \Prx{1+\sum_{k'=1}^K \Pd\eta_{k'}\left|\alpha_{kk'}\right|^2 > 1 + \Pd\sum_{k'\neq k}^K\eta_{k'}\beta_k}\nonumber\\
    &\geq
    \Prx{\Pd\eta_k\left|\alpha_{kk}\right|^2  > \Pd\sum_{k'\neq k}^K\eta_{k'}\beta_k}
    \nonumber\\
    &=
    \Prx{\frac{1}{M}\left\|\B{g}_k\right\|^2 > \frac{1}{M}\sum_{k'\neq
    k}^K\frac{\eta_{k'}}{\eta_k}\beta_k}\nonumber\\
    &
    \to ~ 1, \quad \text{as $M\to\infty$},
\end{align}
where the convergence follows the fact that
$\frac{1}{M}\left\|\B{g}_k\right\|^2 \to \beta_k$ and
$\frac{1}{M}\sum_{k'\neq
    k}^K\frac{\eta_{k'}}{\eta_k}\beta_k \to 0$, as $M\to\infty$.

In addition, by the law of large numbers,
\begin{align}\label{eq:asymp_mr_rayleigh2}
    \frac{
        \left|
            \frac{
                \B{g}_k^H\B{g}_{k'}
                }{
                \left\|\B{g}_{k'}\right\|
                }
        \right|^2
        -\beta_k
            }{
            \left\|\B{g}_k\right\|^2
            }
        &=
        \left(
        \left|
            \frac{
                \B{g}_k^H\B{g}_{k'}
                }{
                M
                }
        \right|^2
            \frac{
                M
                }{
                \left\|\B{g}_{k'}\right\|^2
                }
        - \frac{\beta_k}{M}
        \right)
         \frac{M }{  \left\|\B{g}_k\right\|^2}\nonumber\\
        &\to ~ 0, \quad \text{as $M \to\infty$}.
\end{align}
From \eqref{eq:asymp_mr1}, \eqref{eq:asymp_mr_rayleigh1}, and
\eqref{eq:asymp_mr_rayleigh2}, we obtain
\begin{align}\label{eq:asymp_mr_rayleigh3}
\frac{\hat{\alpha}_{kk}}{\alpha_{kk}}
    \to ~ 1, \quad \text{as $M \to\infty$}.
\end{align}
Our proposed scheme is expected to work very well at large $\tauc$
and $M$.

\item[-] \textbf{Keyhole channels}: Following a similar
methodology used in the case of Rayleigh fading, and using the
identity
\begin{align}\label{eq:proofkh1}
\frac{\B{g}_k^H\B{g}_{k'}}{\|\B{g}_{k'}\|}
    =
    \sqrt{\beta_{k}}\sum_{j=1}^{n_{k}} c_j^{(k)} a_j^{(k)}
    \nu_j^{(k)},
\end{align}
where $\nu_j^{(k)} \triangleq \frac{\left(\B{b}_j^{(k')}
\right)^H\B{g}_{k'}}{\|\B{g}_{k'}\|}$ is $\CG{0}{1}$ distributed,
we can arrive at the same result as \eqref{eq:asymp_mr_rayleigh3}.
The random variable $\nu_j^{(k)}$ is Gaussian due to the fact that
conditioned on $\B{g}_{k'}$, $\nu_j^{(k)}$ is a Gaussian RV with
zero mean and unit variance which is independent of  $\B{g}_{k'}$.

\end{itemize}

\subsubsection{Zero-forcing Processing}

With ZF processing, when $\tauc \to\infty$,
\begin{align}\label{eq:asymp_zf}
\frac{\hat{\alpha}_{kk}}{\alpha_{kk}}
    \to ~ 1, \quad \text{as $M \to\infty$}.
\end{align}
This follows from \eqref{eq:po1as} and the fact that $\alpha_{kk'}
\to 0$, for $k \neq k'$.

\section{Capacity Lower Bound}\label{sec:capbound}
Next, we give a new capacity lower bound for Massive MIMO
with downlink channel gain estimation. It can be applied, in
particular, to our proposed blind channel estimation scheme.\footnote{In Massive MIMO, the bounding technique   in \cite{JAMV:11:WCOM,HBD:13:JSAC} is
commonly used due to its simplicity. This bound is, however, tight only when the effective channel gain $\alpha_{kk}$ hardens. As we show in Section~\ref{sec:CH}, channel hardening does not always hold (for example, not in keyhole channels).  A detailed comparison between our new bound and the bound in \cite{JAMV:11:WCOM,HBD:13:JSAC} is given  in Section~\ref{com:bound}.} Denote by
$\B{y}_k \triangleq [y_k(1) ~ \ldots ~ y_k(\taud)]^T$,
$\B{s}_k\triangleq [s_k(1) ~ \ldots ~ s_k(\taud)]^T$, and
$\B{w}_k\triangleq [w_k(1) ~ \ldots ~ w_k(\taud)]^T$. Then from
\eqref{eq:sys3}, we have
\begin{align}\label{eq:bound1}
\B{y}_k = \sqrt{\Pd\eta_k}\alpha_{kk} \B{s}_k + \sum_{k'\neq k}^K
\sqrt{\Pd\eta_{k'}} \alpha_{kk'}\B{s}_{k'} +
    \B{w}_k.
\end{align}

The capacity of \eqref{eq:bound1} is lower bounded by the mutual
information between the unknown transmitted signal $\B{s}_k$ and
the observed/known values $\B{y}_k$, $\hat{\alpha}_{kk}$. More
precisely, for any distribution of $\B{s}_k$, we obtain the
following capacity bound for the $k$th user:
\begin{align}\label{eq:bound3}
&C_k
    \geq
        \frac{1}{\taud}\I(\B{y}_k,\hat{\alpha}_{kk};\B{s}_k)\nonumber\\
    &=
        \frac{1}{\taud}
        \big[
            \h(\B{s}_k) - \h(\B{s}_k|\B{y}_k,\hat{\alpha}_{kk})
        \big]
    \nonumber\\
    &\!\mathop {=}\limits^{(a)}\!
    \frac{1}{\taud}\h(\B{s}_k) \!-\! \frac{1}{\taud}\!\Big[ \h\big( {s}_k(1)|\B{y}_k,\hat{\alpha}_{kk}\big) \!+\!
    \h \big({s}_k(2)|{s}_k(1),\B{y}_k,\hat{\alpha}_{kk}\big)
    \nonumber\\
    & \hspace{1.2cm}
     + \ldots +  \h\big({s}_k(\taud)|{s}_k(1), \ldots, {s}_k(\taud-1),
    \B{y}_k,\hat{\alpha}_{kk}\big)\Big]\nonumber\\
     &\mathop {\geq}\limits^{(b)}
    \frac{1}{\taud}\h(\B{s}_k)
    - \frac{1}{\taud}\big[\h\left({s}_k(1)|\B{y}_k,\hat{\alpha}_{kk}\right) + \h\left({s}_k(2)|\B{y}_k,\hat{\alpha}_{kk}\right) \nonumber\\&\hspace{3.5cm}+ \ldots +
    \h\left({s}_k(\taud)|\B{y}_k,\hat{\alpha}_{kk}\right)\big],
\end{align}
where in $(a)$ we have used the chain rule \cite{CT:91:Book}, and in
$(b)$ we have used the fact that conditioning reduces entropy.

It is difficult to compute
$\h\left({s}_k(n)|\B{y}_k,\hat{\alpha}_{kk}\right)$ in
\eqref{eq:bound3} since $\hat{\alpha}_{kk}$ and ${s}_k(n)$ are
correlated. To render the problem more tractable, we introduce new
variables $\{\hat{\hat{\alpha}}_{kk}(n)\}$, $n=1, ..., \taud$, which
can be considered as the channel estimates of $\alpha_{kk}$ using
Algorithm~\ref{algo:blce}, but $\xi_k$ is now computed as
\begin{align*}
\frac{|y_k(1)|^2+ \ldots + |y_k(n-1)|^2+ |y_k(n+1)|^2
\ldots+ |y_k(\taud)|^2}{\taud-1}.
\end{align*}
Clearly, $\hat{\hat{\alpha}}_{kk}(n)$ is very close to
$\hat{\alpha}_{kk}$. More importantly,
$\hat{\hat{\alpha}}_{kk}(n)$ is independent of ${s}_{k'}(n)$,
$k'=1, ..., K$. This fact will be used for subsequent derivation
of the capacity lower bound.

Since $\hat{\hat{\alpha}}_{kk}(n)$ is a deterministic function of $\B{y}_k$,
$\h\left({s}_k(n)|\B{y}_k,\hat{\alpha}_{kk}\right)=\h\left({s}_k(n)|\B{y}_k,\hat{\alpha}_{kk},\hat{\hat{\alpha}}_{kk}(n)\right)$,
and hence, \eqref{eq:bound3} becomes
\begin{align}\label{eq:bound3b}
C_k
    &\geq
    \frac{1}{\taud}\h(\B{s}_k)
    - \frac{1}{\taud}\Big[\h\left({s}_k(1)|\B{y}_k,\hat{\alpha}_{kk},\hat{\hat{\alpha}}_{kk}(1)\right)\nonumber\\&\hspace{2.5cm} + \ldots +
    \h\left({s}_k(\taud)|\B{y}_k,\hat{\alpha}_{kk},\hat{\hat{\alpha}}_{kk}(\taud)\right)\Big]\nonumber\\
    &\geq
    \frac{1}{\taud}\h(\B{s}_k)
    - \frac{1}{\taud}\Big[\h\left({s}_k(1)|{y}_k(1),\hat{\hat{\alpha}}_{kk}(1)\right) \nonumber\\&\hspace{1.8cm}+ \ldots +
    \h\left({s}_k(\taud)|{y}_k(\taud),\hat{\hat{\alpha}}_{kk}(\taud)\right)\Big],
\end{align}
where in the last inequality, we have used again the fact that
conditioning reduces entropy. The bound \eqref{eq:bound3b} holds
irrespective of the distribution of $\B{s}_k$. By taking $s_k(1),
\ldots, s_k(\taud)$ to be i.i.d.\ $\CG{0}{1}$, we obtain
\begin{align}\label{eq:bound4a}
C_k
    &\geq
    \log_2(\pi e) -
    \h\left({s}_k(1)|{y}_k(1),\hat{\hat{\alpha}}_{kk}(1)\right).
\end{align}

The right hand side of \eqref{eq:bound4a} is the mutual information
between ${y}_k(1)$ and ${s}_k(1)$ given the side information
$\hat{\hat{\alpha}}_{kk}(1)$. Since $\hat{\hat{\alpha}}_{kk}(1)$ and
${s}_{k'}(1)$, $k'=1, ..., K$, are independent, we have
\begin{align}\label{eq:cond1}
    &\EX{ \left.\bar{w}_k(1) \right| \hat{\hat{\alpha}}_{kk}(1)}
    =
    0,\nonumber\\
    &\EX{ \left.s_{k}^\ast(1)\bar{w}_k(1) \right| \hat{\hat{\alpha}}_{kk}(1)}
    =
    0,\nonumber\\
    &\EX{ \left.\alpha_{kk}^\ast s_{k}^\ast(1)\bar{w}_k(1) \right| \hat{\hat{\alpha}}_{kk}(1)}
    =
    0,
\end{align}
where $\bar{w}_k(1)\triangleq \sum_{k'\neq k}^K \sqrt{\Pd\eta_{k'}}\alpha_{kk'}s_{k'}(1) + w_k(1)$.
Hence we can apply the result in \cite{Mur:00:IT} to further bound
the capacity for the $k$th user as \eqref{eq:bound4}, shown at the top of  the next page.\footnote{The core argument
behind the bound  \eqref{eq:bound4} is the
maximum-entropy property of  Gaussian noise \cite{Mur:00:IT}.
Prompted by a comment from the reviewers, we   stress that
 to obtain \eqref{eq:bound4}, it is not sufficient
    that  the effective noise and the desired signal are uncorrelated.
It is also required that
the  effective noise and the desired signal are
 uncorrelated, \emph{conditioned on the side information.}
}

\setcounter{eqnback}{\value{equation}} \setcounter{equation}{42}
\begin{figure*}[!t]
\begin{align}\label{eq:bound4}
C_k
    &\geq
    R_{k}^{\text{blind}}
    \triangleq
    \EX{
        \log_2
        \left(
        1
        +
        \frac{
            \left|\EX{\left.y_k^\ast(1) s_k(1)\right|\hat{\hat{\alpha}}_{kk}(1) }\right|^2
            }{
            \EX{\left.\left| y_k(1)\right|^2 \right|\hat{\hat{\alpha}}_{kk}(1) } - \left|\EX{\left.y_k^\ast(1) s_k(1)\right|\hat{\hat{\alpha}}_{kk}(1) }\right|^2
            }
        \right)
        },
\end{align}
\hrulefill
\begin{align}\label{eq:bound5}
R_{k}^{\text{blind}}
    =
    \EX{
        \log_2
        \left(
        1
        +
        \frac{
            \Pd\eta_k\left|\EX{\left.\alpha_{kk}\right|\hat{\hat{\alpha}}_{kk}(1) }\right|^2
            }{
            1+\Pd\sum_{k'=1}^K\eta_{k'}\EX{\left.\left| \alpha_{kk'}\right|^2 \right|\hat{\hat{\alpha}}_{kk}(1) }
            - \Pd\eta_k\left|\EX{\left.\alpha_{kk}\right|\hat{\hat{\alpha}}_{kk}(1) }\right|^2
            }
        \right)
        },
\end{align}
\hrulefill \setcounter{equation}{45}
\begin{align}\label{eq:bound5re1}
R_{k}^{\text{blind}}
    \!=\!
    \sum_i p_{\hat{\hat{\alpha}}_{kk}(1)}\!\left(x_i\right )\triangle_{x_i}
     \!\log_2\!\!
        \left(\!\!
        1
        \!+\!
        \frac{
            \Pd\eta_k\left|\EX{\left.\alpha_{kk}\right|x_i }\right|^2
            }{
            1+\Pd\!\!\sum\limits_{k'=1}^K\eta_{k'}\EX{\left.\left| \alpha_{kk'}\right|^2 \right|x_i }
            - \Pd\eta_k\left|\EX{\left.\alpha_{kk}\right|x_i }\right|^2
            }
        \!\!\right),
\end{align}
\hrulefill
\end{figure*}
\setcounter{eqncnt}{\value{equation}}
\setcounter{equation}{\value{eqnback}}

Inserting \eqref{eq:sys3} into \eqref{eq:bound4}, we obtain a capacity
lower bound (achievable rate) for the $k$th user given by \eqref{eq:bound5} at the top of  the next page.

\begin{remark}\label{remark:computing_bound}
  The computation of the capacity lower bound \eqref{eq:bound5}  involves the
  expectations $\EX{\left.\left| \alpha_{kk'}\right|^2
\right|\hat{\hat{\alpha}}_{kk}(1) }$ and $\EX{\left.\alpha_{kk}\right|\hat{\hat{\alpha}}_{kk}(1) }$
 which cannot be directly
computed. However, we can compute $\EX{\left.\left| \alpha_{kk'}\right|^2
\right|\hat{\hat{\alpha}}_{kk}(1) }$
  and $\EX{\left.\alpha_{kk}\right|\hat{\hat{\alpha}}_{kk}(1) }$
 numerically by first using
Bayes's rule and then discretizing it using the Riemann sum:
\setcounter{equation}{44}\begin{align}\label{eq:BR1}
    \EX{X|y}
    &=\int_x x p_{X|Y}(x|y) dx = \int_{x} x \frac{p_{X,Y}(x,y)}{p_{Y}(y)}
    dx \nonumber\\
    &\approx
    \sum_i x_i  \frac{p_{X,Y}(x_i,y)}{p_{Y}(y)} \triangle_{x_i},
\end{align}
where $\triangle_{x_i} \triangleq x_i-x_{i-1}$. Precise steps to
compute \eqref{eq:bound5} are:
\begin{itemize}
    \item[1.] Generate $N$ random realizations of the channel
    $\B{G}$. Then the corresponding $N\times 1$ random vectors of
    $\alpha_{kk}$, $\left| \alpha_{kk'}\right|^2$, and $\hat{\hat{\alpha}}_{kk}(1)$ are obtained.

    \item[2.] From sample vectors obtained in step $1$, numerically
      build the density function
      $\{p_{\hat{\hat{\alpha}}_{kk}(1)}\left(x_i\right)\}$ and the
      joint density functions $\{p_{\alpha_{kk},
        \hat{\hat{\alpha}}_{kk}(1)}\left(y_j,x_i\right)\}$ and
      $p_{\left| \alpha_{kk'}\right|^2,
        \hat{\hat{\alpha}}_{kk}(1)}\left(z_n,x_i \right)$. These
      density functions can be numerically computed using built-in
      functions in MATLAB such as ``$\mathsf{kde}$'' and
      ``$\mathsf{kde2d}$''.

    \item[3.] Using \eqref{eq:BR1}, we compute the achievable rate
    \eqref{eq:bound5} as \eqref{eq:bound5re1}, shown at the top of the next page, where
\setcounter{equation}{46}\begin{align}\label{eq:bound5re2}
&\EX{\left.\alpha_{kk}\right|x_i }
    =
    \sum_j y_j \triangle_{y_j}
    \frac{p_{\alpha_{kk}, \hat{\hat{\alpha}}_{kk}(1)}\left(y_j,x_i\right)}{p_{\hat{\hat{\alpha}}_{kk}(1)}\left(x_i\right)},\\
 &\EX{\!\left.\left| \alpha_{kk'}\right|^2 \right|x_i \!}
   \!=\!
    \sum_n \! z_n \triangle_{z_n}
    \frac{p_{\left| \alpha_{kk'}\right|^2, \hat{\hat{\alpha}}_{kk}(1)}\left(z_n,x_i
    \right)}{p_{\hat{\hat{\alpha}}_{kk}(1)}\left(x_i\right)}.
\end{align}

\end{itemize}

\end{remark}

\begin{remark}
The bound \eqref{eq:bound5} relies on a worst-case Gaussian noise argument \cite{Mur:00:IT}.
Since  the effective noise is the sum of many
random terms, its distribution is, by the central limit theorem,   close to Gaussian.
Hence, our bounds are expected to be rather tight and   they are likely to closely represent
what state-of-the-art coding would deliver in reality.  (This is generally true
for the capacity lower bounds used in much of the Massive MIMO literature; see for example,
  quantitative examples
  in \cite[Myth~4]{BLM:16:CM}.)
\end{remark}

\section{Numerical Results and Discussions}\label{sec:numerical-result}

In this section, we provide numerical results to evaluate our proposed
channel estimation scheme. We consider the per-user normalized MSE and
net throughput as performance metrics. We define
$$\mathsf{SNR}_{\text{d}} = \Pd\times \text{median[cell-edge large-scale fading]},$$ and
$$\mathsf{SNR}_{\text{u}} = \Pu\times \text{median[cell-edge large-scale fading]},$$ where the cell-edge large-scale fading is the large-scale fading between the BS and a user located at the cell-edge.
This gives $\mathsf{SNR}_{\text{d}}$ and $\mathsf{SNR}_{\text{u}}$
the interpretation of the median downlink and the uplink cell-edge
SNRs. For keyhole channels, we assume $n_k=n_{\tt KH}$ and
$c_{j}^{(k)} = 1/\sqrt{n_{\tt KH}}$, for all $k=1, \ldots, K$ and
$j=1, \ldots, n_{\tt KH}$.

In all examples, we compare the performances of three cases: i) ``use
$\EX{\alpha_{kk}}$'', representing the case when the $k$th user relies
on the statistical properties of the channels, i.e., it uses
$\EX{\alpha_{kk}}$ as estimate of $\alpha_{kk}$; ii) ``DL pilots
\cite{NLM:13:ACCCC}'', representing the use of beamforming training
\cite{NLM:13:ACCCC} with linear MMSE channel estimation; and iii)
``proposed scheme'', representing our proposed downlink blind channel
estimation scheme (using Algorithm~1). In our proposed scheme, the
curves with $\taud=\infty$ correspond to the case that the $k$th user
perfectly knows the asymptotic value of $\xi_k$. Furthermore, we
choose $\taup=K$. For the beamforming training scheme, the duration of
the downlink training is chosen as $\taudp=K$.

\subsection{Normalized Mean-Square Error}

\begin{figure}[t!]
\centering
\subfigure[Rayleigh fading channels]{%
\includegraphics[width=0.45\textwidth]{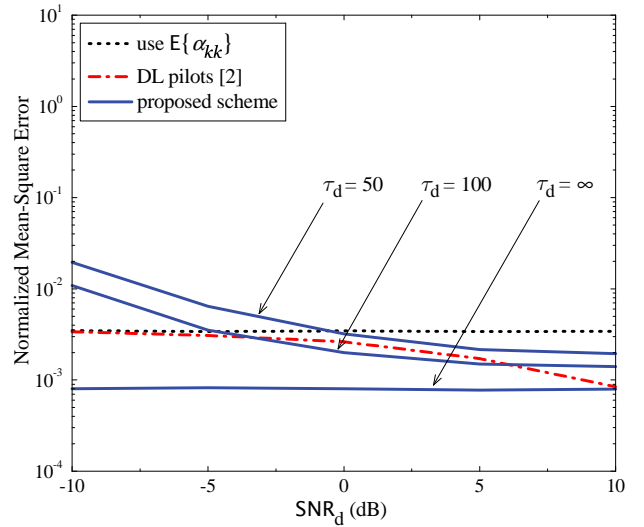}
\label{fig:MSEvsSNR_mr_rayleigh}} \quad
\subfigure[Single-keyhole channels]{%
\includegraphics[width=0.45\textwidth]{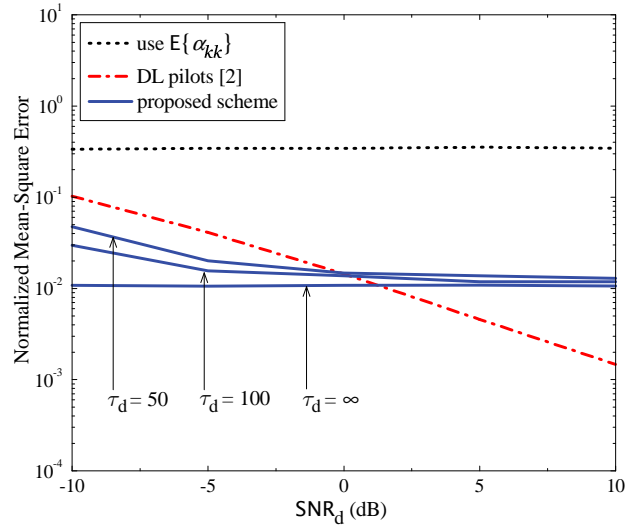}
\label{fig:MSEvsSNR_mr_keyhole}} \caption{Normalized MSE versus
$\mathsf{SNR}_{\text{d}}$ for different channel estimation
schemes, for MR processing. Here, $M=100$, $K=10$, and
$\mathsf{SNR}_{\text{u}} = 0$~dB.} \label{fig:MSEvsSNR_mr}
\end{figure}

\begin{figure}[t!]
\centering
\subfigure[Rayleigh fading channels]{%
\includegraphics[width=0.45\textwidth]{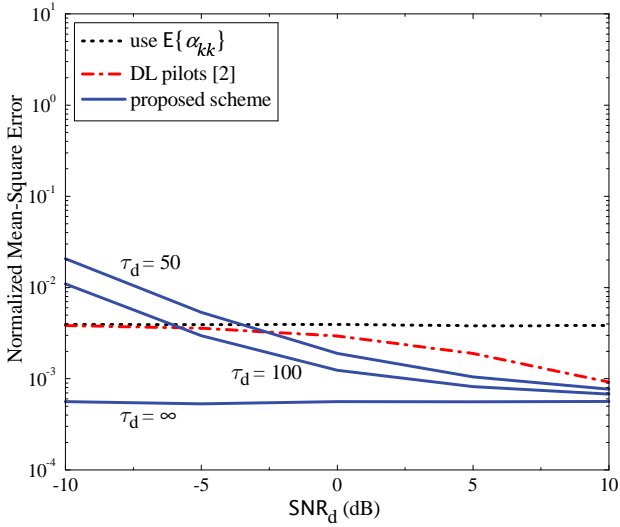}
\label{fig:MSEvsSNR_zf_rayleigh}} \quad
\subfigure[Single-keyhole channels]{%
\includegraphics[width=0.45\textwidth]{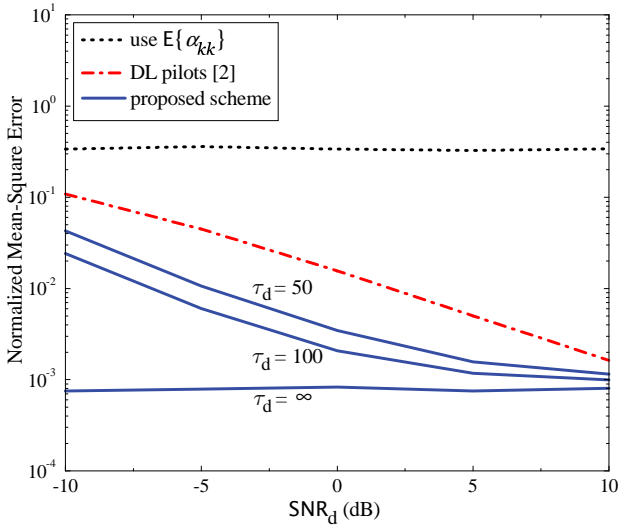}
\label{fig:MSEvsSNR_zf_keyhole}} \caption{Same as
Figure~\ref{fig:MSEvsSNR_mr}, but for ZF processing.}
\label{fig:MSEvsSNR_zf}
\end{figure}

We consider the normalized MSE at user $k$, defined as:
\begin{align}\label{eq:NMSE1a}
{\tt MSE}_k \triangleq
\frac{\EX{\left|\hat{\alpha}_{kk}-\alpha_{kk}
\right|^2}}{\left|\EX{\alpha_{kk}} \right|^2}.
\end{align}
In this part, we choose $\beta_k=1$, and equal power allocation to
all users, i.e, $\eta_k=1/K$, $\forall k$.
Figures~\ref{fig:MSEvsSNR_mr} and \ref{fig:MSEvsSNR_zf} show the
normalized MSE versus $\mathsf{SNR}_{\text{d}}$ for MR
and ZF processing, respectively, under Rayleigh fading
and single-keyhole channels. Here, we choose $M=100$, $K=10$, and
$\mathsf{SNR}_{\text{u}}=0$~dB.

We can see that, in Rayleigh fading channels, for both
MR and ZF processing, the MSEs of the three
schemes (use $\EX{\alpha_{kk}}$, DL pilots, and proposed scheme)
are comparable. Using $\EX{\alpha_{kk}}$ in lieu of the true
$\alpha_{kk}$ for signal detection works rather well. However, in
keyhole channels, since the channels do not harden, the MSE when
using $\EX{\alpha_{kk}}$ as the estimate of $\alpha_{kk}$ is very
large. In both propagation environments, our proposed scheme works
very well and improves when $\taud$ increases (since the
approximation in \eqref{eq:appro1} becomes tighter). Our scheme
outperforms the beamforming training scheme for a wide range of
SNRs, even for short coherence intervals. The training-based method uses the received pilot signals  only  during
 a short time,
to estimate the effective channel gain. In contrast,  our proposed scheme uses the received data during a whole coherence block. This is the basic reason for  why  our proposed  scheme can perform better than the training-based scheme. (Note also that the training-based method is
based on linear MMSE estimation,  which is suboptimal, but that is a second-order effect.)


\begin{figure}[t!]
\centerline{\includegraphics[width=0.47\textwidth]{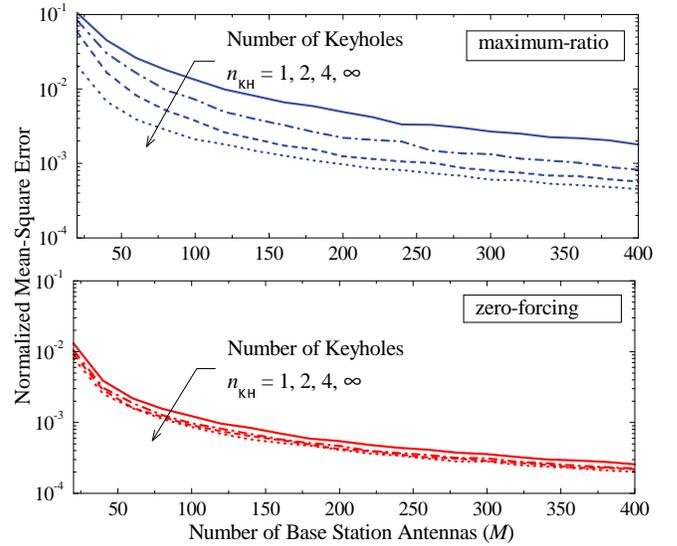}}
\caption{Normalized MSE versus $M$ for different number of
keyholes $n_k=n_{\tt KH}$, using Algorithm~\ref{algo:blce}. Here,
$\mathsf{SNR}_{\text{u}}=0$~dB, $\mathsf{SNR}_{\text{d}}=5$~dB,
and $K=10$. \label{fig:MSE_M_keyhole}}
\end{figure}

Next we study the affects of the number of BS antennas and the
number of keyholes on the performance of our proposed scheme. We
choose $K=10$, $\taud=100$, $\mathsf{SNR}_{\text{u}}=0$~dB, and
$\mathsf{SNR}_{\text{d}}=5$~dB. Figure~\ref{fig:MSE_M_keyhole}
shows the normalized MSE versus $M$ for different numbers of
keyholes $n_{\tt KH}$ with MR and ZF processing. When $n_{\tt
KH}=\infty$, we have Rayleigh fading. As expected, the MSE reduces
when $M$ increases. More importantly, our proposed scheme works
well even when $M$ is not large. Furthermore, we can see that the
MSE does not change much when the number of keyholes varies. This
implies the robustness of our proposed scheme against the
different propagation environments.

Note that, with the beamforming training scheme in
\cite{NLM:13:ACCCC}, we additionally have to spend at least $K$
symbols on training pilots (this is not accounted for here, since we
only evaluate MSE). By contrast, our proposed scheme does not require
any resources for downlink training. To account for the loss due to
training, we will examine the net throughput in the next part.

\subsection{Downlink Net Throughput}\label{sec:num_throughput}

The downlink net throughputs of three cases---use
$\EX{\alpha_{kk}}$, DL pilots, and proposed schemes---are defined
as:
\begin{align}\label{eq:nrSE1a}
\mathcal{S}_k^{\text{noCSI}}
    &=
    B \frac{\taud}{\tauc} R_k^{\text{noCSI}},\\\label{eq:nrSE1b}
\mathcal{S}_k^{\text{pilot}}
    &=
    B \frac{\taud-\taudp}{\tauc} R_k^{\text{pilot}},\\\label{eq:nrSE1c}
\mathcal{S}_k^{\text{blind}}
    &=
    B \frac{\taud}{\tauc} R_k^{\text{blind}},
\end{align}
where $B$ is the spectral bandwidth, $\tauc$ is again the
coherence interval in symbols, and $\taud$ is the number of
symbols per coherence interval allocated for downlink
transmission. Note that $R_k^{\text{noCSI}}$,
$R_k^{\text{pilot}}$, and $R_k^{\text{blind}}$  are the
corresponding achievable rates of these cases.
$R_k^{\text{blind}}$ is given by \eqref{eq:bound5}, while
$R_k^{\text{pilot}}$ and $R_k^{\text{noCSI}}$ can be computed by
using \eqref{eq:bound5}, but $\hat{\hat{\alpha}}_{kk}(1)$ is
replaced with the channel estimate of $\alpha_{kk}$ using scheme
in \cite{NLM:13:ACCCC} respectively $\EX{\alpha_{kk}}$. The term
$\dfrac{\taud}{\tauc}$ in \eqref{eq:nrSE1a} and \eqref{eq:nrSE1c}
comes from the fact that, for each coherence interval of $\tauc$
samples, with our proposed scheme and the case of no channel
estimation, we spend $\taud$ samples for downlink payload data
transmission. The term $\dfrac{\taud-\taudp}{\tauc}$ in
\eqref{eq:nrSE1b} comes from the fact that we spend
$\taudp$ symbols on downlink pilots to estimate the
effect channel gains \cite{NLM:13:ACCCC}. In all examples, we
choose $B=20$~MHz and $\taud=\tauc/2$ (half of the coherence
interval is used for downlink transmission).

\begin{figure}[t!]
\centering
\subfigure[Rayleigh fading channels]{%
\includegraphics[width=0.45\textwidth]{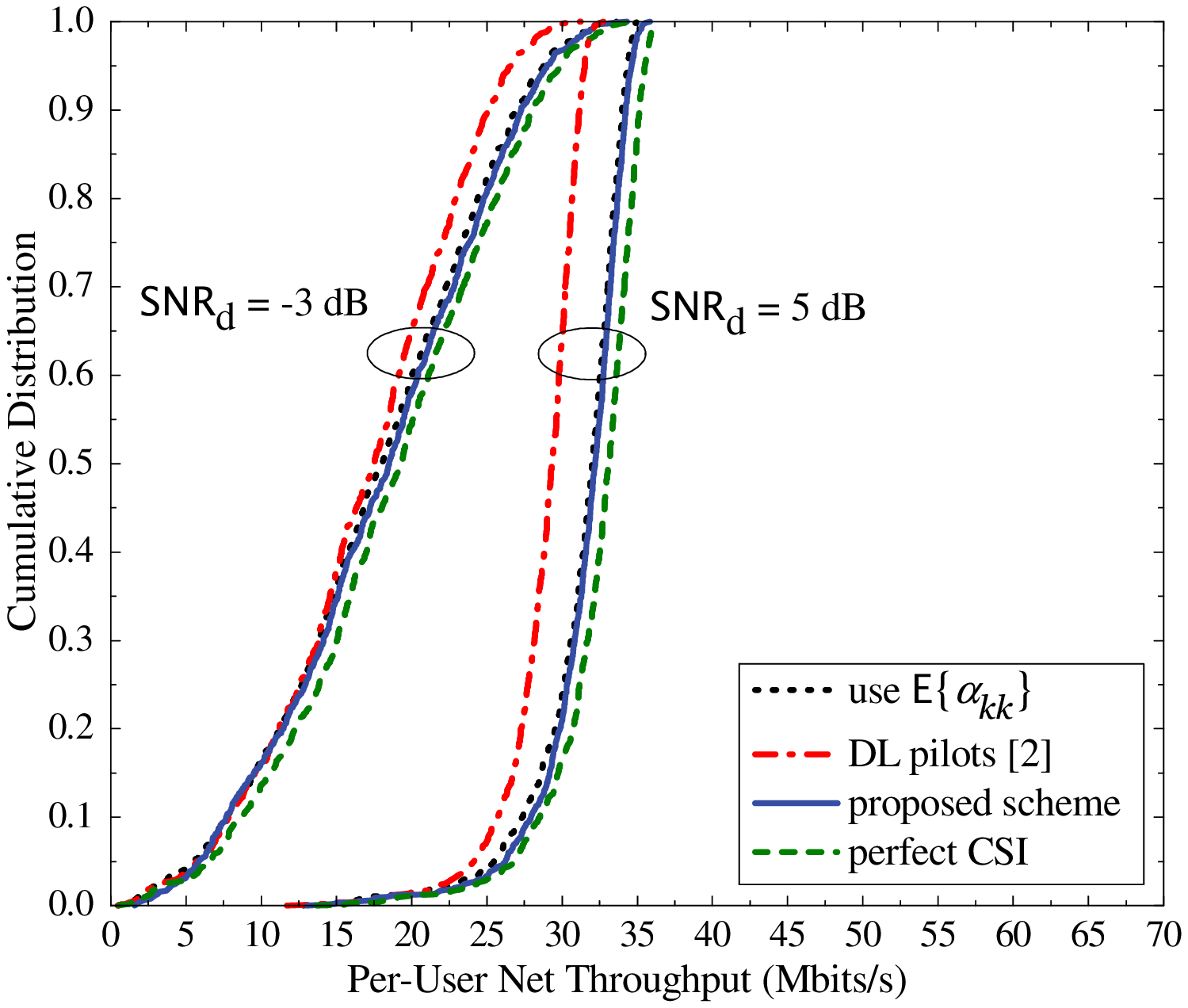}
\label{fig:cdf_mr_rayleigh}} \quad
\subfigure[Single-keyhole channels]{%
\includegraphics[width=0.45\textwidth]{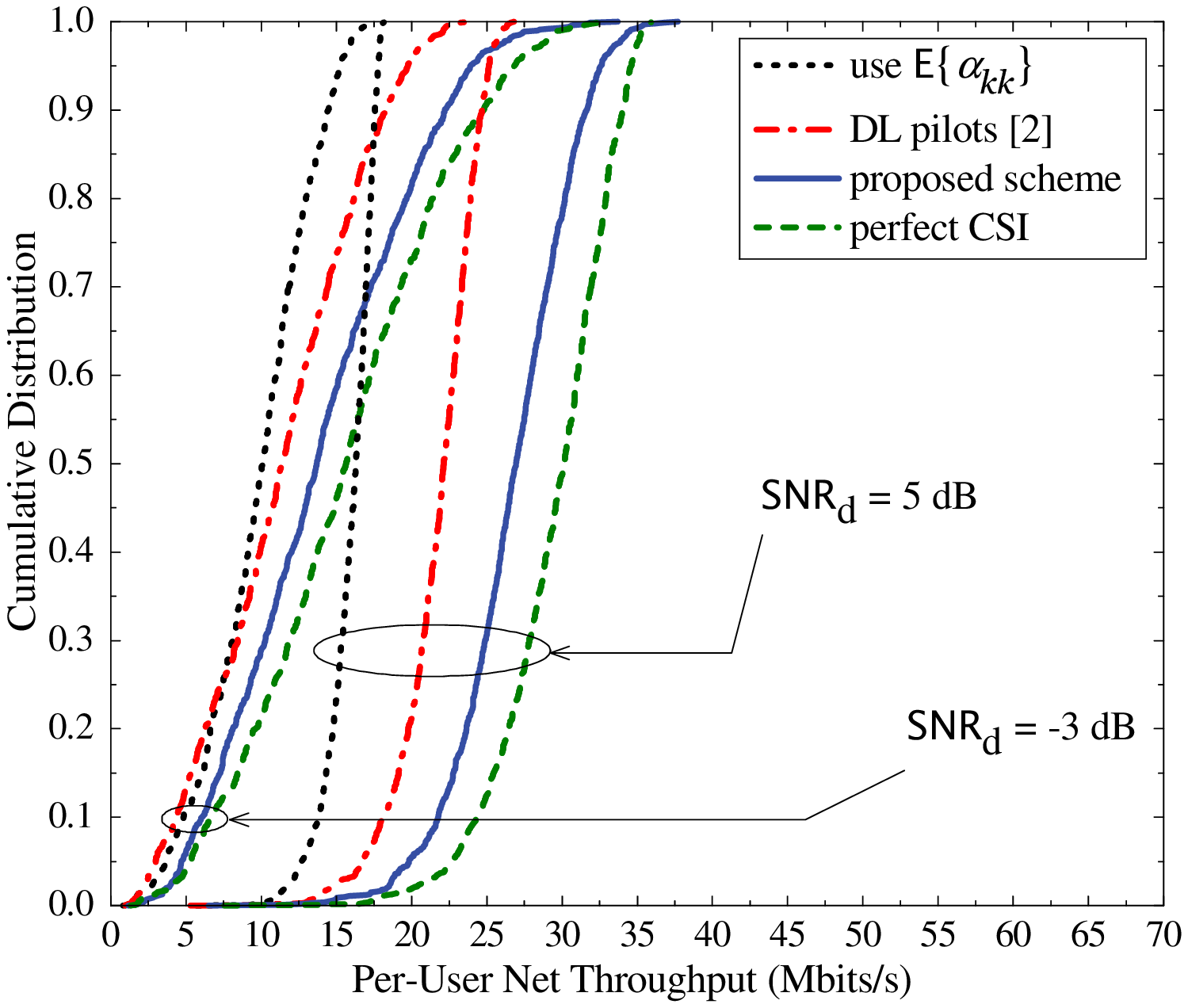}
\label{fig:cdf_mr_keyhole}} \caption{The cumulative distribution
of the per-user downlink net throughput for MR
processing. Here, $M=100$, $K=10$, $\tauc=200$ ($\taud=100$),
$\mathsf{SNR}_{\text{d}}=10\mathsf{SNR}_{\text{u}}$, and
$B=20$~MHz.} \label{fig:cdf_mr}
\end{figure}

\begin{figure}[t!]
\centering
\subfigure[Rayleigh fading channels]{%
\includegraphics[width=0.45\textwidth]{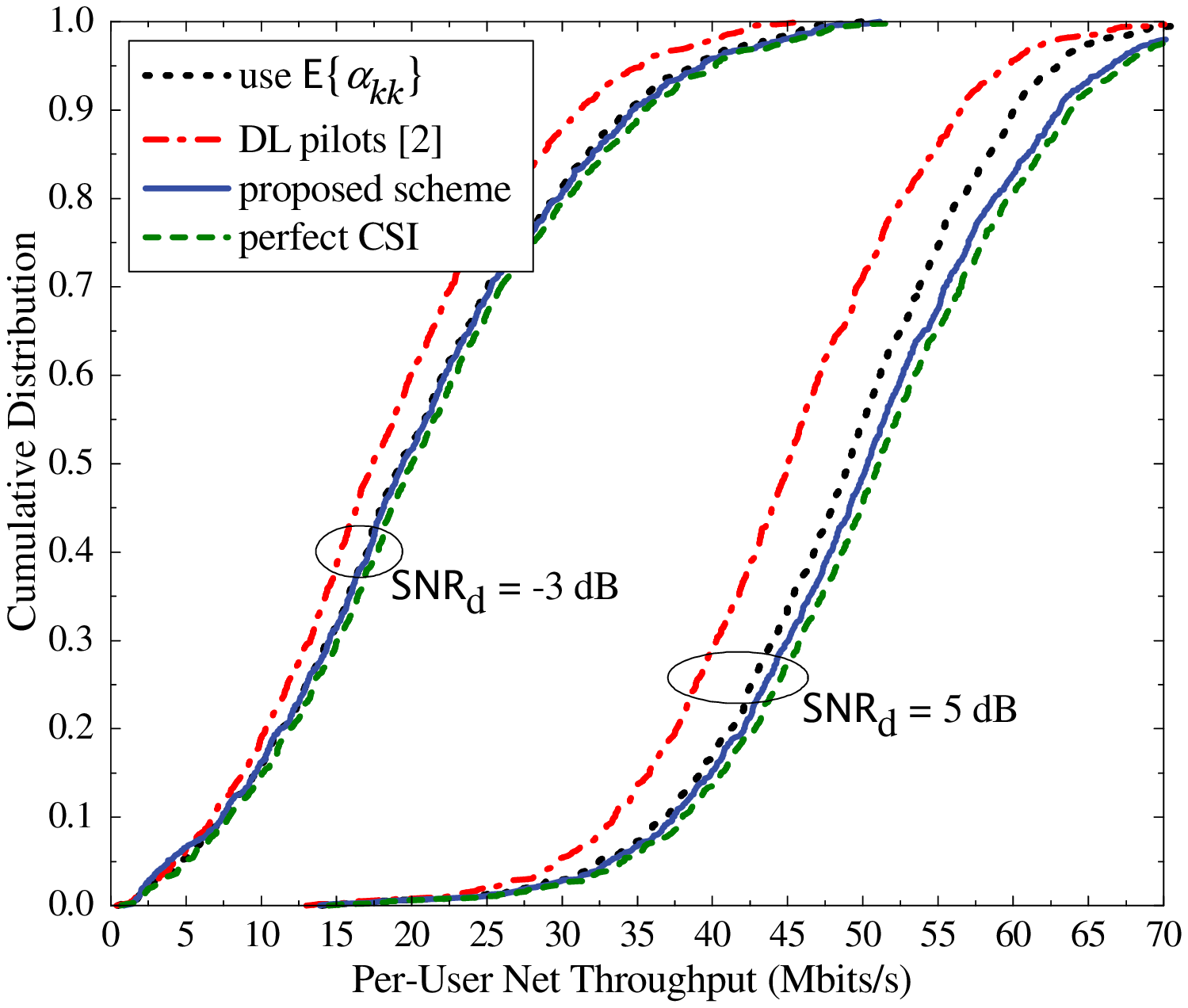}
\label{fig:cdf_zf_rayleigh}} \quad
\subfigure[Single-keyhole channels]{%
\includegraphics[width=0.45\textwidth]{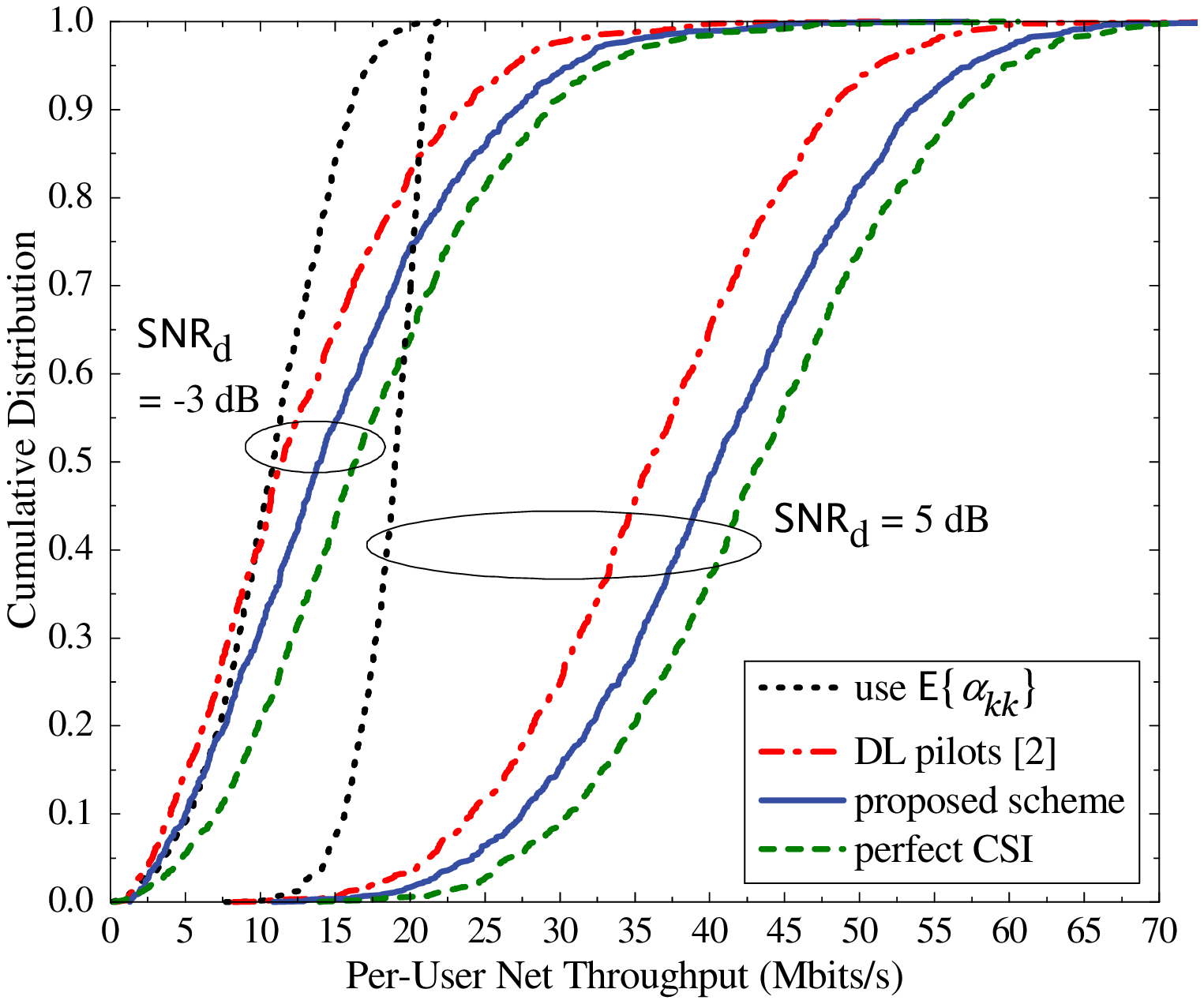}
\label{fig:cdf_zf_keyhole}} \caption{Same as
Figure~\ref{fig:cdf_mr}, but for ZF processing.}
\label{fig:cdf_zf}
\end{figure}

We consider a more realistic scenario which incorporates the
large-scale fading and max-min power control:
\begin{itemize}
\item To generate the large-scale fading, we consider an
  annulus-shaped cell with a radius of $R_{\text{max}}$ meters, and
  the BS is located at the cell center.  $K+1$ users are
  placed uniformly at random in the cell with a minimum distance of
  $R_{\text{min}}$ meters from the BS. The user with the
  smallest large-scale fading $\beta_k$ is dropped, such that $K$
  users remain. The large-scale fading is modeled by path loss,
  shadowing (with log-normal distribution), and random user locations:
\begin{align}\label{eq:lare-scale}
    \beta_k = \text{PL}_0\left(\frac{d_k}{R_{\text{min}}}\right)^\upsilon \times
    10^{\frac{\sigma_{\text{sh}}\cdot \mathcal{N}(0,1)}{10}},
\end{align}
where $\upsilon$ is the path loss exponent and $\sigma_{\text{sh}}$ is
the standard deviation of the shadow fading. The factor $\text{PL}_0$
in \eqref{eq:lare-scale} is a reference path loss constant which is
chosen to satisfy a given downlink cell-edge SNR,
$\mathsf{SNR}_{\text{d}}$. In the simulation, we choose
$R_{\text{min}}=100$, $R_{\text{max}}=1000$, $\upsilon=3.8$, and
$\sigma_{\text{sh}}=8$~dB. We generate $1000$ random realizations of
user locations and shadowing fading profiles.

    \item The power control control coefficients $\{\eta_k\}$ are
      chosen from the max-min power control algorithm
      \cite{YM:14:VTC}:
\begin{align}\label{eq:pccoeffs}
    \eta_k = \left\{\begin{array}{ll}
      \frac{1+\Pd\beta_k
                }{
                \Pd\gamma_k\left(\frac{1}{ \Pd}\sum\limits_{k'=1}^K\frac{1}{ \gamma_{k'}}+\sum\limits_{k'=1}^K\frac{\beta_{k'}}{ \gamma_{k'}}\right)}, & \text{for MR,} \\
      \frac{1+\Pd(\beta_k-\gamma_k)
                }{
                \Pd\gamma_k\left(\frac{1}{\Pd}\sum\limits_{k'=1}^K\frac{1}{ \gamma_{k'}}+\sum\limits_{k'=1}^K\frac{\beta_{k'}-\gamma_{k'}}{
                \gamma_{k'}}\right)},
                 & \text{for ZF.} \\
    \end{array}\right.
\end{align}
     This max-min
    power control offers uniformly good service for all users for the case where the $k$th user uses $\EX{\alpha_{kk}}$ as estimate of $\alpha_{kk}$.

\end{itemize}

\begin{figure}[t!]
\centering
\subfigure[Rayleigh fading channels]{%
\includegraphics[width=0.45\textwidth]{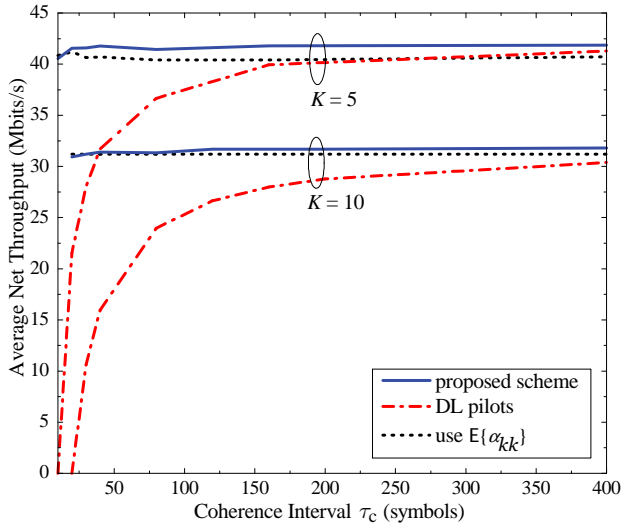}
\label{fig:ave_mr_rayleigh}} \quad
\subfigure[Single-keyhole channels]{%
\includegraphics[width=0.45\textwidth]{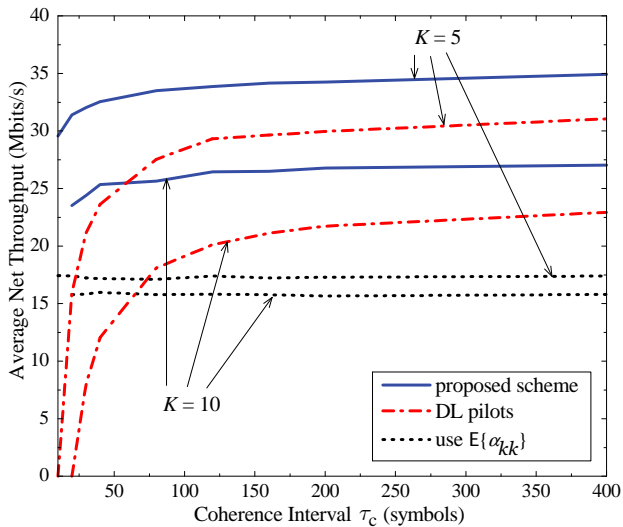}
\label{fig:ave_mr_keyhole}} \caption{The average per-user downlink net throughput for MR processing. Here, $M=100$,
$\mathsf{SNR}_{\text{d}}=10\mathsf{SNR}_{\text{u}}= 5$~dB, and
$B=20$~MHz.} \label{fig:average_mr}
\end{figure}

Figures~\ref{fig:cdf_mr} and \ref{fig:cdf_zf} show the cumulative
distributions of the per-user downlink net throughput for
MR respectively ZF processing, under Rayleigh
fading and single-keyhole channels. Here we choose $M=100$,
$K=10$, $\tauc=200$, and $\mathsf{SNR}_{\text{d}}=10
\mathsf{SNR}_{\text{u}}$. As a baseline for comparisons, we
additionally add the curves labelled ``perfect CSI''. These curves
represent the presence of a genie receiver
at the $k$th user, which knows the channel gain  perfectly. For both
propagation environments, our proposed scheme is the best and
performs very close to the genie receiver.  For Rayleigh
fading channels, due to the hardening property of the channels,
our proposed scheme and the scheme using statistical property of
the channels are comparable. These schemes perform better than the
beamforming training scheme in \cite{NLM:13:ACCCC}. The reason is
that, with beamforming training scheme, we have to spend $\taudp$
pilot samples for the downlink training. For single-keyhole
channels, the channels do not harden, and hence, it is necessary
to estimate the effective channel gains. Our proposed scheme
improves the system performance significantly. At
$\mathsf{SNR}_{\text{d}}=5$~dB, with MR processing, our
proposed scheme can improve the 95\%-likely net throughput by
about $20$\% and $60$\%, compared with the downlink beamforming
training scheme respectively the case of without channel
estimation. With ZF processing, our proposed scheme can
improve the 95\%-likely net throughput by $15$\% and $66$\%,
respectively. The  MSE of ``use $\EX{\alpha_{kk}}$'' does not depend on $\mathsf{SNR}_{\text{d}}$ (see Figures~\ref{fig:MSEvsSNR_mr} and \ref{fig:MSEvsSNR_zf}), but it depends on $\mathsf{SNR}_{\text{u}}$. In Figures~5 and 6, when $\mathsf{SNR}_{\text{d}}$ increases,  $\mathsf{SNR}_{\text{u}}$ also increases, and hence, the per-user throughput gaps between the ``use $\EX{\alpha_{kk}}$'' curves and the ``perfect CSI'' curves  vary as  $\mathsf{SNR}_{\text{d}}$ increases.

 Finally, we investigate the effect of the coherence interval $\tauc$ and the number of users $K$ on the performance of our proposed scheme. Figure~\ref{fig:average_mr} shows the average downlink net throughput versus $\tauc$ with MR processing for different $K$ in both Rayleigh fading and keyhole channels. The average is taken over the large-scale fading. Our proposed scheme overcomes the disadvantage of beamforming training scheme in high mobility environments  (short coherence interval), and the disadvantage of statistical property-based scheme in non-hardening propagation environments, and hence, performs very well in many cases, even when $\tauc$ and $K$ are small.

\begin{figure}[t!]
\centering
\subfigure[Rayleigh fading channels]{%
\includegraphics[width=0.45\textwidth]{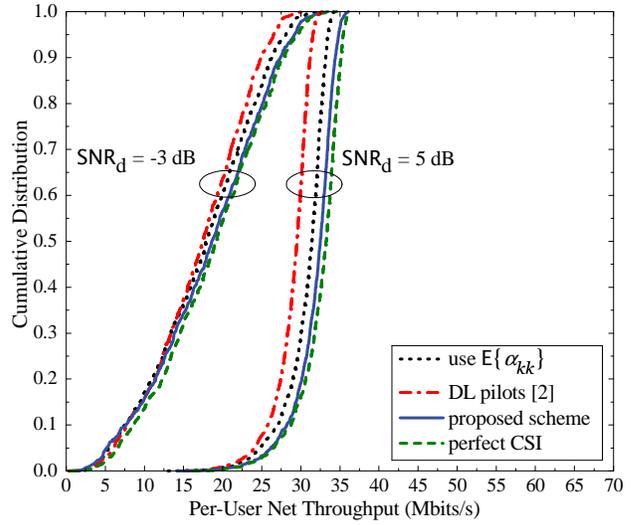}
\label{fig:cdf_mr_rayleigh_long}} \quad
\subfigure[Single-keyhole channels]{%
\includegraphics[width=0.45\textwidth]{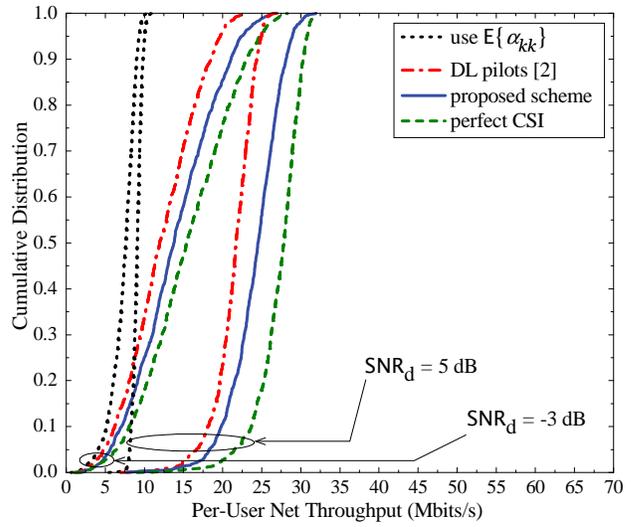}
\label{fig:cdf_mr_keyhole_long}} \caption{Same as
Figure~\ref{fig:cdf_mr}, but with long-term average power
constraint \eqref{eq:mrlong1}.} \label{fig:cdf_mr_long}
\end{figure}

\section{Comments}\label{sec:comments}

\subsection{Short-Term V.s. Long-Term Average Power Constraint}

The precoding vectors $\B{a}_k$ in \eqref{eq:mr1} and \eqref{eq:zf111}
are chosen to satisfy a short-term average power constraint where the
expectation of \eqref{eq:powerconst} is taken over only
$\B{s}(n)$. This short-term average power constraint is not the only
possibility. Alternatively, one could consider a long-term average
power constraint where the expectation in \eqref{eq:powerconst} is
taken over $\B{s}(n)$ and over the small-scale fading. With
MR combining, the long-term-average-power-based precoding
vectors $\{\B{a}_k\}$ are
\begin{align}\label{eq:mrlong1}
    \B{a}_k
    =
    \frac{\hat{\B{g}}_k}{\sqrt{\EX{\|\hat{\B{g}}_k\|^2}}}
    =
    \frac{\hat{\B{g}}_k}{\sqrt{M\gamma_k}}, \quad k=1,
\ldots, K.
\end{align}

However, with ZF, the long-term-average-power-based precoder
is not always valid. For example, for single-keyhole channels,
perfect uplink estimation, and $K=1$, we have
\begin{align}\label{eq:zflong1}
    \EX{\left\|\left[\B{G}\left(\B{G}^H\B{G}\right)^{-1}\right]_k \right\|^2}
,
\end{align}
which is infinite.

We emphasize here that compared to the short-term average power case,
the long-term average power case does not make a difference in the
sense that the resulting effective channel gain does not always
harden, and hence, it needs to be estimated.  (The harding property of
the channels is discussed in detail in Section~\ref{sec:CH}.) To see
this more quantitatively, we compare the performance of three cases:
``use $\EX{\alpha_{kk}}$'', ``DL pilots \cite{NLM:13:ACCCC}'', and
``proposed scheme'' for MR with long-term average power
constraint \eqref{eq:mrlong1}. As seen in
Figure~\ref{fig:cdf_mr_long}, under keyhole channels, our proposed
scheme improves the net throughput significantly, compared to the
``use $\EX{\alpha_{kk}}$'' case.

\subsection{Flaw of the Bound in \cite{NLM:13:ACCCC,ZZYJL:15:VT}}

In the above numerical results, the curves with downlink pilots are obtained
by first replacing $\hat{\hat{\alpha}}_{kk}(1)$ in \eqref{eq:bound5}
with the channel estimate obtained using the algorithm in
\cite{NLM:13:ACCCC}, and then using the numerical technique discussed
in Remark~\ref{remark:computing_bound} to compute the capacity bound.

Closed-form expressions for achievable rates with downlink training
were given in \cite[Eq.~(12)]{NLM:13:ACCCC} and \cite{ZZYJL:15:VT}.
However, those formulas were not rigorously correct, since
$\{a_{kk'}\}$ are non-Gaussian in general (even in Rayleigh fading)
and hence the linear MMSE estimate is not equal to the MMSE estimate;
the expressions for the capacity bounds in
\cite{NLM:13:ACCCC,ZZYJL:15:VT} are valid only when the MMSE estimate
is inserted. However, the expressions \cite{NLM:13:ACCCC,ZZYJL:15:VT}
are likely to be extremely accurate approximations.  A similar
approximation was stated in \cite{KM:15:COM}.

\begin{figure}[t!]
\centering
\subfigure[Rayleigh fading channels]{%
\includegraphics[width=0.45\textwidth]{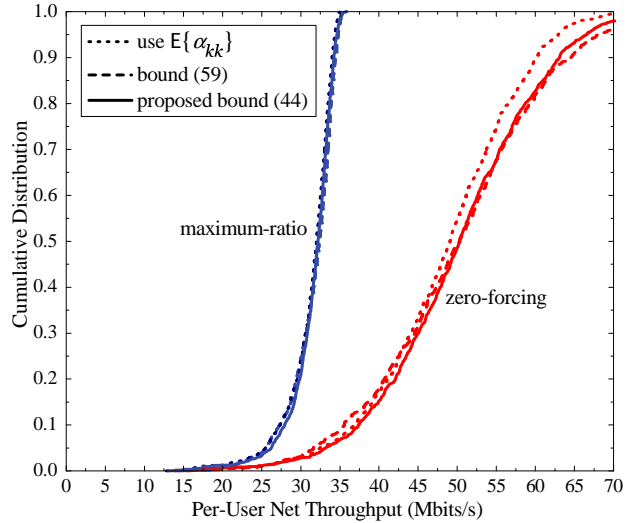}
\label{fig:bound_compare_rayleigh}} \quad
\subfigure[Single-keyhole channels]{%
\includegraphics[width=0.45\textwidth]{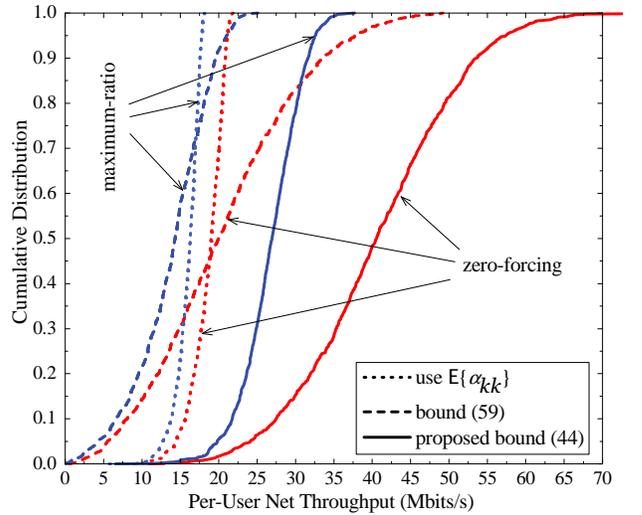}
\label{fig:bound_compare_keyhole}} \caption{The cumulative distribution of the per-user downlink net throughput for MR  and ZF processing. Here, $M=100$, $K=10$, $\tauc=200$ ($\taud=100$),
$\mathsf{SNR}_{\text{d}}=10\mathsf{SNR}_{\text{u}} =5$ dB, and
$B=20$~MHz.} \label{fig:bound_compare}
\end{figure}
\subsection{Using the Capacity  Bounding Technique of  \cite{JAMV:11:WCOM,HBD:13:JSAC}}\label{com:bound}

It may be tempting to use   the bounding technique in \cite{JAMV:11:WCOM,HBD:13:JSAC} to derive a simpler capacity bound
as follows (the index $n$ is omitted for simplicity of notation):
\begin{itemize}
 \item[i)] Divide the received signal $\eqref{eq:sys3}$ by the channel estimate $\hat{\hat{\alpha}}_{kk}$,
 \begin{align}\label{eq:boundco1}
y'_k
&=
\frac{y_k}{\sqrt{\Pd\eta_k}\hat{\hat{\alpha}}_{kk}}\nonumber\\
     &=
        \frac{\alpha_{kk}}{\hat{\hat{\alpha}}_{kk}} s_k + \sum_{k'\neq k}^K \sqrt{\frac{\eta_{k'}}{\eta_k}}\frac{\alpha_{kk'}}{\hat{\hat{\alpha}}_{kk}}s_{k'} +
    \frac{w_k}{\sqrt{\Pd\eta_k}\hat{\hat{\alpha}}_{kk}}.
\end{align}

\item[ii)] Rewrite \eqref{eq:boundco1} as the sum of the desired signal multiplied
with a deterministic gain, $\EX{\frac{\alpha_{kk}}{\hat{\hat{\alpha}}_{kk}}} s_k$, and  remaining terms which constitute uncorrelated   effective noise,
 \begin{align}\label{eq:boundco2}
y'_k
&=
        \EX{\frac{\alpha_{kk}}{\hat{\hat{\alpha}}_{kk}}} s_k + \left(\frac{\alpha_{kk}}{\hat{\hat{\alpha}}_{kk}}-\EX{\frac{\alpha_{kk}}{\hat{\hat{\alpha}}_{kk}}}\right) s_k\nonumber\\ &+ \sum_{k'\neq k}^K \sqrt{\frac{\eta_{k'}}{\eta_k}}\frac{\alpha_{kk'}}{\hat{\hat{\alpha}}_{kk}}s_{k'} +
    \frac{w_k}{\sqrt{\Pd\eta_k}\hat{\hat{\alpha}}_{kk}}.
\end{align}
\end{itemize}

\setcounter{eqnback}{\value{equation}} \setcounter{equation}{58}
\begin{figure*}[!t]
 \begin{align}\label{eq:boundco3}
 R_k^{\text{UnF}}
    =
    \log_2\left(
    1 + \frac{\left|\EX{\frac{\alpha_{kk}}{\hat{\hat{\alpha}}_{kk}}}\right|^2 }{\varx{\frac{\alpha_{kk}}{\hat{\hat{\alpha}}_{kk}} } + \sum\limits_{k'\neq k}^K \frac{\eta_{k'}}{\eta_k} \EX{\left|\frac{\alpha_{kk'}}{\hat{\hat{\alpha}}_{kk}}\right|^2} + \frac{1}{\Pd\eta_k}\EX{\frac{1}{\left|\hat{\hat{\alpha}}_{kk}\right|^2}} }   \right).
\end{align}
\hrulefill
\end{figure*}
\setcounter{eqncnt}{\value{equation}}
\setcounter{equation}{\value{eqnback}}

The  worst-case Gaussian noise property   \cite{Mur:00:IT} then yields the capacity bound \eqref{eq:boundco3}, shown at the top of the next page. This bound does not require the complicated numerical computation
 given in Section~\ref{sec:capbound}. However, this bound is tight only when the effective channel gain $\alpha_{kk}$ hardens, which is generally not the case
 under   the models that we consider herein.

 More quantitatively, Figure~\ref{fig:bound_compare} shows a comparison between
  our new bound \eqref{eq:bound5} and the bound \eqref{eq:boundco3}.
  The figure shows   the cumulative
distributions of the per-user downlink net throughput for
MR and ZF processing, for the same setup  as in Section~\ref{sec:num_throughput}. In Rayleigh fading, the throughputs for the three cases ``use
$\EX{\alpha_{kk}}$'', ``bound \eqref{eq:boundco3}'', and ``proposed bound \eqref{eq:bound5}'', are very close, and hence, relying on statistical channel knowledge ($\EX{\alpha_{kk}})$ for signal detection is good enough -- obviating the
need for the bound in \eqref{eq:boundco3}. In in keyhole channels, the bound \eqref{eq:boundco3} is significantly inferior to our proposed bound. Therefore,
the bound \eqref{eq:boundco3}
is of no use neither in Rayleigh fading nor in keyhole channels.

\section{Conclusion} \label{sec:conclusion}

In the Massive MIMO downlink, in propagation environments where
the channel hardens, using the mean of the effective channel gain
for signal detection is good enough. However, the channels may not
always harden. Then, to reliably decode the transmitted signals,
each user should estimate its effective channel gain rather than
approximate it by its mean. We proposed a new blind channel
estimation scheme at the users which does not require any downlink
pilots. With this scheme, the users can blindly estimate their
effective channel gains directly from the data received during a
coherence interval. Our proposed channel estimation scheme is
computationally easy, and performs very well.  Numerical results
show that  in non-hardening propagation environments and for large
numbers of BS antennas, our proposed scheme significantly
outperforms both the downlink beamforming training scheme in
\cite{NLM:13:ACCCC} and the conventional approach that
approximates the effective channel gains by their means.


\appendix

\subsection{Derivation  of \eqref{eq:ds 5b}}\label{sec:AppeKU1}
\setcounter{equation}{59}
We have,
\begin{align}\label{eq:proofds 5b1}
&\frac{\varx{\snorm{\B{g}_k}}
    }{\left(\EX{\snorm{\B{g}_k}}\right)^2}
    \!=\!
    \frac{1}{\beta_k^2M^2}\EX{\left\| \B{g}_k\right\|^4} - \frac{1}{\beta_k^2 M^2}\!\left(\!\EX{\left\|
    \B{g}_k\right\|^2}\right)^2\nonumber\\
    &=
    \frac{1}{\beta_k^2 M^2}\EX{\left\| \B{g}_k\right\|^4} - 1
    \nonumber\\
    &=
    \frac{1}{M^2}\EX{
        \left|
        \sum_{i=1}^{n_k} \sum_{n=1}^{n_k}
        \left( c_i^{(k)} a_i^{(k)}
        \B{b}_i^{(k)}\right)^H
         c_n^{(k)} a_n^{(k)}
        \B{b}_n^{(k)}
        \right|^2
        } -1
    \nonumber\\
    &=
    \frac{1}{M^2}\EX{
        \left|
        \sum_{i=1}^{n_k}\!
        \left\| \tilde{\B{b}}_i^{(k)}\right\|^2
        \!+\!
        \sum_{i=1}^{n_k}\! \sum_{n\neq i}^{n_k}\!
        \left(\tilde{\B{b}}_i^{(k)}\right)^H\!
         \tilde{\B{b}}_n^{(k)}
        \right|^2
        } \!-\!1,
\end{align}
where $ \tilde{\B{b}}_i^{(k)}\triangleq c_i^{(k)} a_i^{(k)}
\B{b}_i^{(k)}$. We can see that, the terms in the double sum have
zero mean. We now consider the covariance
        between two arbitrary terms: $$\EX{\left(\tilde{\B{b}}_i^{(k)}\right)^H
         \tilde{\B{b}}_n^{(k)}  \left( \left(\tilde{\B{b}}_{i'}^{(k)}\right)^H
         \tilde{\B{b}}_{n'}^{(k)} \right)^\ast},$$
where $i\neq n$, $i'\neq n'$, and $(i,n)\neq (i',n')$. Clearly, if
$(i,n)\neq (n',i')$, then
$$\EX{\left(\tilde{\B{b}}_i^{(k)}\right)^H
         \tilde{\B{b}}_n^{(k)}  \left( \left(\tilde{\B{b}}_{i'}^{(k)}\right)^H
         \tilde{\B{b}}_{n'}^{(k)} \right)^\ast} = 0.$$ If $(i,n)=
         (n',i')$, the we have
\begin{align}\label{eq:proofds 5b2}
&\EX{\left(\tilde{\B{b}}_i^{(k)}\right)^H
         \tilde{\B{b}}_n^{(k)}  \left( \left(\tilde{\B{b}}_{i'}^{(k)}\right)^H
         \tilde{\B{b}}_{n'}^{(k)} \right)^\ast}
         \nonumber\\&=
         \EX{\left(\tilde{\B{b}}_i^{(k)}\right)^H
         \tilde{\B{b}}_n^{(k)}  \left(\tilde{\B{b}}_{n}^{(k)}\right)^T
         \left(\tilde{\B{b}}_{i}^{(k)}\right)^\ast } = 0,
\end{align}
where we used the fact that if $z$ is a circularly symmetric
complex Gaussian random variable with zero mean, then
$\EX{z^2}=0$. The above result implies that the terms
$\left(\tilde{\B{b}}_i^{(k)}\right)^H
         \tilde{\B{b}}_n^{(k)}$ [inside the double sum
of \eqref{eq:proofds 5b1}] are zero-mean mutual uncorrelated
random variables. Furthermore, they are uncorrelated with
$\sum_{i=1}^{n_k}
        \left\| \tilde{\B{b}}_i^{(k)}\right\|^2$, so \eqref{eq:proofds
        5b1} can be rewritten as:
\begin{align}\label{eq:proofds 5b3}
\frac{\varx{\snorm{\B{g}_k}}
    }{\left(\EX{\snorm{\B{g}_k}}\right)^2}
    &=
    \frac{1}{M^2}\underbrace{\EX{
        \left|
        \sum_{i=1}^{n_k}
        \left\| \tilde{\B{b}}_i^{(k)}\right\|^2
        \right|^2
        }}_{\triangleq \text{Term1}}\nonumber\\
        &\hspace{-1cm}+
    \frac{1}{M^2}\sum_{i=1}^{n_k} \sum_{n\neq i}^{n_k}\underbrace{\EX{
        \left|
        \left(\tilde{\B{b}}_i^{(k)}\right)^H
         \tilde{\B{b}}_n^{(k)}
        \right|^2
        }}_{\triangleq \text{Term2}}
         -1.
\end{align}
We have,
\begin{align}\label{eq:proofds 5b4}
\text{Term1}
    &=
    \sum_{i=1}^{n_k} \EX{ \left\| \tilde{\B{b}}_i^{(k)}\right\|^4 }
    +
    \sum_{i=1}^{n_k}\sum_{n\neq i}^{n_k} \EX{\left\| \tilde{\B{b}}_i^{(k)}\right\|^2 \left\| \tilde{\B{b}}_n^{(k)}\right\|^2}
    \nonumber\\
    &=
    \sum_{i=1}^{n_k} \EX{\left| c_i^{(k)}\right|^4 \left| a_i^{(k)}\right|^4 \left\| {\B{b}}_i^{(k)}\right\|^4 }
    \nonumber\\
    &\hspace{-1cm}+
    \sum_{i=1}^{n_k}\!\sum_{n\neq i}^{n_k}\! \EX{\!\left| c_i^{(k)}\!\right|^2 \!\left| a_i^{(k)}\right|^2 \left\| {\B{b}}_i^{(k)}\right\|^2\!
    }\!\EX{\!\left| c_n^{(k)}\!\right|^2 \left| a_n^{(k)}\right|^2 \!\left\| {\B{b}}_n^{(k)}\right\|^2 \!}
    \nonumber\\
    &=
    2M(M+1)\sum_{i=1}^{n_k} \left| c_i^{(k)}\right|^4
    +
    M^2\sum_{i=1}^{n_k}\sum_{n\neq i}^{n_k} \left| c_i^{(k)}\right|^2
    \left| c_n^{(k)}\right|^2\nonumber\\
    &= M(M+2)\sum_{i=1}^{n_k} \left| c_i^{(k)}\right|^4 + M^2,
\end{align}
where we have used the identity that if $\B{z}\sim
\CG{\mathbf{0}}{\B{I}_n}$, then $\EX{\left\|\B{z}\right\|^4} =
n(n+1)$.

Furthermore, we have
\begin{align}\label{eq:proofds 5b5}
&\text{Term2}
    =
        \EX{
        \left|
        \left(c_i^{(k)} a_i^{(k)} \B{b}_i^{(k)}\right)^H
         c_n^{(k)} a_n^{(k)}\B{b}_n^{(k)}
        \right|^2
        }
    \nonumber\\
    &=
        \left| c_i^{(k)} \right|^2\left| c_n^{(k)} \right|^2
        \E{\left| a_i^{(k)} \right|^2} \EX{\left| a_n^{(k)} \right|^2}
        \E{\left|
        \left(\B{b}_i^{(k)}\right)^H \B{b}_n^{(k)}
        \right|^2
        }\nonumber\\
    &= M \left| c_i^{(k)} \right|^2\left| c_n^{(k)} \right|^2.
\end{align}
Substituting \eqref{eq:proofds 5b4} and \eqref{eq:proofds 5b5}
into \eqref{eq:proofds 5b3}, we obtain
\begin{align}\label{eq:proofds 5b6}
\frac{\varx{\snorm{\B{g}_k}}
    }{\left(\EX{\snorm{\B{g}_k}}\right)^2}
    &=
    \left(1+\frac{1}{M}\right) \sum_{i=1}^{n_k} \left|
    c_i^{(k)}\right|^4 + \frac{1}{M}.
\end{align}

\subsection{Derivation  of \eqref{eq:mean_interference1}} \label{sec:app-mean_interference1}

Here, we provide the proof  of \eqref{eq:mean_interference1}.
\begin{itemize}

\item With MR, for both Rayleigh and
keyhole channels, $\B{g}_k$ and $\B{a}_{k'}$ are
    independent, for $k\neq k'$. Thus,     we have
\begin{align}
 \EX{|\alpha_{kk'}|^2}
    &=
    \EX{\B{a}_{k'}^H\B{g}_k\B{g}_{k}^H \B{a}_{k'}}\nonumber\\
    &=
    \beta_k\EX{\left\|\B{a}_{k'}\right\|^2}\nonumber\\
    &=\beta_k.
\end{align}

\item With ZF, for Rayleigh channels, the channel estimate
$\hat{\B{g}}_k$ is independent of the channel estimation error
$\tilde{\B{g}}_k$. So $\tilde{\B{g}}_k$ and $\B{a}_{k'}$ are
independent. In addition, from \eqref{eq:zf111}, we have
$$\hat{\B{g}}_{k}^H \B{a}_{k'} = 0, \quad k\neq k',$$ and therefore,
\begin{align}
 \EX{|\alpha_{kk'}|^2}
    &=
    \EX{|\B{g}_{k}^H \B{a}_{k'}|^2}\nonumber\\
    &=
    \EX{|\tilde{\B{g}}_{k}^H \B{a}_{k'}|^2} \nonumber\\
    &=
    \EX{\B{a}_{k'}^H\tilde{\B{g}}_{k}\tilde{\B{g}}_{k}^H \B{a}_{k'}} \nonumber\\
    &=
    (\beta_k-\gamma_k)\EX{\left\|\B{a}_{k'}\right\|^2} \nonumber\\
    &=\beta_k-\gamma_k.
\end{align}

\end{itemize}


\begin{IEEEbiography}
{Hien Quoc Ngo}  received the B.S. degree in electrical engineering from Ho Chi Minh City University of Technology, Vietnam, in 2007. He then received the M.S. degree in Electronics and Radio Engineering from Kyung Hee University, Korea, in 2010, and the Ph.D. degree in communication systems from Link\"oping University (LiU), Sweden, in 2015. From May to December 2014, he visited Bell Laboratories, Murray Hill, New Jersey, USA.

Hien Quoc Ngo is currently a postdoctoral researcher of the Division for Communication Systems in the Department of Electrical Engineering (ISY) at Link\"oping University, Sweden. He is also a Visiting Research Fellow at the School of Electronics, Electrical Engineering and Computer Science, Queen's University Belfast, U.K. His current research interests include massive (large-scale) MIMO systems and cooperative communications.

Dr. Hien Quoc Ngo received the IEEE ComSoc Stephen O. Rice Prize in Communications Theory in 2015. He also received the IEEE Sweden VT-COM-IT Joint Chapter Best Student Journal Paper Award in 2015. He was an \emph{IEEE Communications Letters} exemplary reviewer for 2014, an \emph{IEEE Transactions on Communications} exemplary reviewer for 2015. He has been a member of Technical Program Committees for several IEEE conferences such as ICC, Globecom, WCNC, VTC, WCSP, ISWCS, ATC, ComManTel.
\end{IEEEbiography}

\begin{IEEEbiography}
{Erik G. Larsson} is Professor of Communication Systems
at Link\"oping University (LiU) in Link\"oping, Sweden. He
previously worked for the Royal Institute of
Technology (KTH) in Stockholm, Sweden, the
University of Florida, USA, the George Washington University, USA,
and Ericsson Research, Sweden.  In 2015 he was
a Visiting Fellow at Princeton University, USA, for four months. He received his Ph.D. degree from Uppsala University,
Sweden, in 2002.

His main professional interests are within the areas of wireless
communications and signal processing. He has co-authored some 130 journal papers
on these topics, he is co-author of the two Cambridge University Press textbooks \emph{Space-Time
Block Coding for Wireless Communications} (2003) and \emph{Fundamentals of Massive MIMO} (2016).
 He is co-inventor on 16 issued and many pending patents on wireless technology.

He served as Associate Editor for, among others, the \emph{IEEE Transactions on
Communications} (2010-2014) and \emph{IEEE Transactions on Signal Processing} (2006-2010).
He serves as  chair of the IEEE Signal Processing Society SPCOM technical committee in 2015--2016 and
he served as chair of the steering committee for the \emph{IEEE Wireless
Communications Letters} in 2014--2015.  He was the General Chair of the Asilomar Conference
on Signals, Systems and Computers in 2015, and Technical Chair in
2012.

He received the \emph{IEEE Signal Processing Magazine} Best Column Award twice, in 2012 and 2014, and
 the IEEE ComSoc Stephen O. Rice Prize in Communications Theory in 2015.
  He is a Fellow of the IEEE.
\end{IEEEbiography}


\begin{thebibliography}{10}
\providecommand{\url}[1]{#1}
\csname url@samestyle\endcsname
\providecommand{\newblock}{\relax}
\providecommand{\bibinfo}[2]{#2}
\providecommand{\BIBentrySTDinterwordspacing}{\spaceskip=0pt\relax}
\providecommand{\BIBentryALTinterwordstretchfactor}{4}
\providecommand{\BIBentryALTinterwordspacing}{\spaceskip=\fontdimen2\font plus
\BIBentryALTinterwordstretchfactor\fontdimen3\font minus
  \fontdimen4\font\relax}
\providecommand{\BIBforeignlanguage}[2]{{%
\expandafter\ifx\csname l@#1\endcsname\relax
\typeout{** WARNING: IEEEtran.bst: No hyphenation pattern has been}%
\typeout{** loaded for the language `#1'. Using the pattern for}%
\typeout{** the default language instead.}%
\else
\language=\csname l@#1\endcsname
\fi
#2}}
\providecommand{\BIBdecl}{\relax}
\BIBdecl

\bibitem{NL:15:ICASSP}
H.~Q. Ngo and E.~G. Larsson, ``Blind estimation of effective downlink channel
  gains in massive {MIMO},'' in \emph{Proc. IEEE International Conference on
  Acoustics, Speech and Signal Processing (ICASSP)}, Brisbane, Australia, Apr.
  2015.

\bibitem{NLM:13:ACCCC}
H.~Q. Ngo, E.~G. Larsson, and T.~L. Marzetta, ``Massive {MU-MIMO} downlink
  {TDD} systems with linear precoding and downlink pilots,'' in \emph{Proc.
  Allerton Conference on Communication, Control, and Computing},
  Urbana-Champaign, Illinois, Oct. 2013.

\bibitem{Eri:13:MCOM}
E.~G. Larsson, F.~Tufvesson, O.~Edfors, and T.~L. Marzetta, ``Massive {MIMO}
  for next generation wireless systems,'' \emph{{IEEE} Commun. Mag.}, vol.~52,
  no.~2, pp. 186--195, Feb. 2014.

\bibitem{LLSAZ:14:JSTSP}
L.~Lu, G.~Y. Li, A.~L. Swindlehurst, A.~Ashikhmin, and R.~Zhang, ``An overview
  of massive {MIMO}: Benefits and challenges,'' \emph{IEEE J.\ Select.\ Topics
  Signal Process.}, vol.~8, no.~5, pp. 742--758, Oct. 2014.

\bibitem{BL:16:VTM}
T.~Bogale and L.~B. Le, ``Massive {MIMO} and {mmWave} for {5G} wireless
  {HetNet}: Potentials and challenges,'' \emph{IEEE Veh.\ Technol.\ Mag.},
  vol.~11, no.~1, pp. 64--75, Feb. 2016.

\bibitem{GERT:15:WCOM}
X.~Gao, O.~Edfors, F.~Rusek, and F.~Tufvesson, ``Massive {MIMO} performance
  evaluation based on measured propagation data,'' \emph{{IEEE} Trans. Wireless
  Commun.}, vol.~14, no.~7, pp. 3899--3911, Jul. 2015.

\bibitem{ZJWZM:14:SSP}
Q.~Zhang, S.~Jin, K.-K. Wong, H.~Zhu, and M.~Matthaiou, ``Power scaling of
  uplink massive {MIMO} systems with arbitrary-rank channel means,'' \emph{IEEE
  J.\ Select.\ Topics Signal Process.}, vol.~8, no.~5, pp. 966--981, Oct. 2014.

\bibitem{JAMV:11:WCOM}
J.~Jose, A.~Ashikhmin, T.~L. Marzetta, and S.~Vishwanath, ``Pilot contamination
  and precoding in multi-cell {TDD} systems,'' \emph{{IEEE} Trans. Wireless
  Commun.}, vol.~10, no.~8, pp. 2640--2651, Aug. 2011.

\bibitem{YM:13:JSAC}
H.~Yang and T.~L. Marzetta, ``Performance of conjugate and zero-forcing
  beamforming in large-scale antenna systems,'' \emph{{IEEE} J. Sel. Areas
  Commun.}, vol.~31, no.~2, pp. 172--179, Feb. 2013.

\bibitem{HBD:13:JSAC}
J.~Hoydis, {S. ten Brink}, and M.~Debbah, ``Massive {MIMO} in the {UL/DL} of
  cellular networks: How many antennas do we need?'' \emph{{IEEE} J. Sel. Areas
  Commun.}, vol.~31, no.~2, pp. 160--171, Feb. 2013.

\bibitem{ZZYJL:15:VT}
J.~Zuo, J.~Zhang, C.~Yuen, W.~Jiang, and W.~Luo, ``Multi-cell multi-user
  massive {MIMO} transmission with downlink training and pilot contamination
  precoding,'' \emph{{IEEE} Trans. Veh. Technol.}, vol.~65, no.~8, pp.
  6301--6314, Aug. 2016.

\bibitem{KM:15:COM}
A.~Khansefid and H.~Minn, ``Achievable downlink rates of {MRC} and {ZF}
  precoders in massive {MIMO} with uplink and downlink pilot contamination,''
  \emph{{IEEE} Trans. Commun.}, vol.~63, no.~12, pp. 4849--4864, Dec. 2015.

\bibitem{KMC:15:ICC}
T.~Kim, K.~Min, and S.~Choi, ``Study on effect of training for downlink massive
  {MIMO} systems with outdated channel,'' in \emph{Proc. IEEE International
  Conference on Communications (ICC)}, London, UK, Jun. 2015.

\bibitem{NLM:13:TCOM}
H.~Q. Ngo, E.~G. Larsson, and T.~L. Marzetta, ``Energy and spectral efficiency
  of very large multiuser {MIMO} systems,'' \emph{{IEEE} Trans. Commun.},
  vol.~61, no.~4, pp. 1436--1449, Apr. 2013.

\bibitem{VT:03:IT}
P.~Viswanath and D.~N.~C. Tse, ``Sum capacity of the vector {G}aussian
  broadcast channel and uplink-downlink duality,'' \emph{{IEEE} Trans. Inf.
  Theory}, vol.~49, no.~8, pp. 1912--1921, Aug. 2003.

\bibitem{NLM:14:Eusipco}
H.~Q. Ngo, E.~G. Larsson, and T.~L. Marzetta, ``Aspects of favorable
  propagation in massive {MIMO},'' in \emph{Proc. European Signal Processing
  Conf. (EUSIPCO)}, Lisbon, Portugal, Sep. 2014.

\bibitem{LCC:15:WCOM}
Y.-G. Lim, C.-B. Chae, and G.~Caire, ``Performance analysis of massive {MIMO}
  for cell-boundary users,'' \emph{{IEEE} Trans. Wireless Commun.}, vol.~14,
  no.~12, pp. 6827--6842, Dec. 2015.

\bibitem{moustakas2000communication}
A.~L. Moustakas, H.~U. Baranger, L.~Balents, A.~M. Sengupta, and S.~H. Simon,
  ``Communication through a diffusive medium: Coherence and capacity,''
  \emph{Science}, vol. 287, no. 5451, pp. 287--290, 2000.

\bibitem{TV:04:FTCIT}
A.~M. Tulino and S.~Verd\'{u}, ``Random matrix theory and wireless
  communications,'' \emph{Foundations and Trends in Communications and
  Information Theory}, vol.~1, no.~1, pp. 1--182, Jun. 2004.

\bibitem{GBGP:02:COM}
D.~Gesbert, H.~B{\"o}lcskei, D.~A. Gore, and A.~J. Paulraj, ``Outdoor {MIMO}
  wireless channels: Models and performance prediction,'' \emph{{IEEE} Trans.
  Commun.}, vol.~50, no.~12, pp. 1926--1934, Dec. 2002.

\bibitem{ATM:06:WCOM}
P.~Almers, F.~Tufvensson, and A.~F. Molisch, ``Keyhole effect in {MIMO}
  wireless channels: {Measurements} and theory,'' \emph{{IEEE} Trans. Wireless
  Commun.}, vol.~5, no.~12, pp. 3596--3604, Dec. 2006.

\bibitem{LJGM:10:SP}
X.~Li, S.~Jin, X.~Gao, and M.~R. McKay, ``Capacity bounds and low complexity
  transceiver design for double-scattering {MIMO} multiple access channels,''
  \emph{{IEEE} Trans. Signal Process.}, vol.~58, no.~5, pp. 2809--2822, May
  2010.

\bibitem{ZJWM:11:WCOM}
C.~Zhong, S.~Jin, K.-K. Wong, and M.~R. McKay, ``Ergodic mutual information
  analysis for multi-keyhole {MIMO} channels,'' \emph{{IEEE} Trans. Wireless
  Commun.}, vol.~10, no.~6, pp. 1754--1763, Jun. 2011.

\bibitem{LL:IT:11}
G.~Levin and S.~Loyka, ``From multi-keyholes to measure of correlation and
  power imbalance in {MIMO} channels: Outage capacity analysis,'' \emph{{IEEE}
  Trans. Inf. Theory}, vol.~57, no.~6, pp. 3515--3529, Jun. 2011.

\bibitem{CT:91:Book}
T.~M. Cover and J.~A. Thomas, \emph{Elements of Information Theory}.\hskip 1em
  plus 0.5em minus 0.4em\relax New York: Wiley, 1991.

\bibitem{Mur:00:IT}
M.~M\'{e}dard, ``The effect upon channel capacity in wireless communications of
  perfect and imperfect knowledge of the channel,'' \emph{{IEEE} Trans. Inf.
  Theory}, vol.~46, no.~3, pp. 933--946, May 2000.

\bibitem{BLM:16:CM}
E.~Bj{\"{o}}rnson, E.~G. Larsson, and T.~L. Marzetta, ``Massive {MIMO}: 10
  myths and one critical question,'' \emph{{IEEE} Commun. Mag.}, vol.~54,
  no.~2, pp. 114--123, Feb. 2016.

\bibitem{YM:14:VTC}
H.~Yang and T.~L. Marzetta, ``A macro cellular wireless network with uniformly
  high user throughputs,'' in \emph{Proc. IEEE Veh. Technol. Conf. (VTC)}, Sep.
  2014.

\end{thebibliography}
\end{document}